\documentclass[sort&compress]{elsarticle}

\usepackage{rotating}
\usepackage{amsmath}
\usepackage{amssymb}
\usepackage{algorithmicx}
\usepackage{algorithm2e}
\usepackage{hyperref}

\usepackage{graphicx}
\usepackage{subcaption}
\newcommand{\IR}{\mathbb{R}}
\newcommand{\FF}{\textsc{f}}
\renewcommand{\SS}{\textsc{s}}
\newcommand{\fin}{\textsc{fin}}
\newcommand{\bs}[1]{\boldsymbol{#1}}

\newtheorem{remark}{Remark}

\address{Eindhoven University of Technology, PO Box 513, 5600 MB Eindhoven, The Netherlands}
\begin{document}

\begin{frontmatter}
\title{On the robustness of Dirichlet--Neumann coupling schemes for fluid-structure-interaction problems with nearly-closed fluid domains}

\author[tue,pav]{A. Aissa--Berraies}
\author[pav]{F.A. Auricchio}
\author[eva]{G.J. van Zwieten}
\author[tue]{E.H.\ van\ Brummelen\corref{cor}}

\address[tue]{Multiscale Engineering Fluid Dynamics group, Department of Mechanical Engineering, Eindhoven University of Technology, The Netherlands}
\address[pav]{Computational Mechanics and Advanced Materials Group, Department of Civil Engineering and Architecture, University of Pavia, Italy}
\address[eva]{Evalf Computing, Delft, The Netherlands}
\cortext[cor]{Corresponding author}

\begin{abstract}%
The partitioned approach for fluid-structure interaction (FSI) simulations involves solving the structural and flow field problems sequentially. This approach allows for separate settings for the fluid and solid subsystems and hence modularity, thereby leveraging advanced commercial and open-source software capabilities to offer increased flexibility for diverse FSI applications. Most partitioned FSI schemes apply the Dirichlet--Neumann partitioning of 
the interface conditions. The Dirichlet--Neumann coupling scheme has proven adequate in a wide range of applications. However, this coupling scheme is sensitive to the added-mass effect, and it is susceptible to the incompressibility dilemma, i.e. it completely fails for FSI problems in which the fluid is 
incompressible and furnished with Dirichlet boundary conditions on the part of its boundary complementary to the interface. In the present paper, we demonstrate that if the fluid is incompressible and the fluid domain is nearly-closed, in the sense that the fluid domain is furnished with Dirichlet conditions except for a permeable part of the boundary where a Robin-type condition holds, then the Dirichlet--Neumann partitioned approach is sensitive to the flow resistance at the permeable part and, in particular, convergence of the partitioned approach deteriorates as the flow resistance increases.
The Dirichlet--Neumann partitioned approach then becomes arbitrarily unstable in the limit of vanishing permeability, i.e., if the flow resistance passes to infinity. Based on a simple model problem, we establish that in the nearly-closed case, the convergence rate of the Dirichlet--Neumann partitioned method depends on a so-called {\it added-damping effect\/}. The presented analysis provides insights that can be leveraged to improve the robustness and efficiency of partitioned approaches for FSI problems involving contact, such as valve opening/closing applications. In addition, the results elucidate the incompressibility dilemma as a formal limit of the added-damping effect passing to infinity, and the corresponding challenges related to FSI problems with nearly closed fluid-domain configurations. Based on numerical experiments, we consider the generalization of the results of the simple model problem to more complex nearly-closed FSI problems.
\end{abstract}


\end{frontmatter}

%
\section{Introduction}
\label{introduction}
Fluid-structure interaction (FSI) is becoming increasingly important in a variety of engineering and physical disciplines. In recent years, numerical FSI methods have developed into important tools for industrial decision-making on complex production technologies supported by simulation-based engineering. The importance of FSI simulation methods lies in the wide range of multiphysics applications they can be used for, including, for instance, predicting the aeroelastic stability of aircraft~\cite{farhat2004cfd,Membrane-wing}, assessing the fluid's effect on valves during operation~\cite{Ogawa1995,Lin2020,Liu2019,natarajan2017analysis,Aissa-Berraies:2025fv}, ensuring the safety of vehicle passengers via airbag deployment\cite{timo_van2015} \cite{timo_thesis}, and investigating cardiovascular disorders \cite{Radtke2017,cai2019some}.

Computational simulation of a fluid flow interacting with a deforming solid can be performed using two types of approaches, viz.  monolithic or partitioned. In monolithic methods, the fluid and solid subsystems are combined and treated concurrently in a unified framework, i.e. the subsystems are, in principle, formulated as a single system of nonlinear equations, which is in turn solved by means of a solution strategy that disregards the composite structure of the system. Monolithic approaches have been advocated for their robustness and stability~\cite{Michler2005}, which renders them suitable for strongly coupled and non-conventional FSI problems, such as elasto-capillarity~\cite{Brummelen:2021wt}. On the other hand, monolithic methods generally require considerable code customization and a significant effort to complete the setup with respect to a particular problem specification. Furthermore, in many cases, the generated system matrices are excessively
large~\cite{chimakurthi2018ansys} and, in addition, on account of the aggregation of subsystems with distinct properties, the matrices are typically severely ill-conditioned~\cite{Richter:2015kq}. As a result, monolithic approaches have so far been mostly restricted to in-house codes in academia, and have not found widespread application in state-of-the-art commercial and open-source software packages, which are primarily used to address complex industrial applications. As opposed to monolithic methods, partitioned approaches are based on solving the flow and structure subproblems separately~\cite{farhat2004cfd,van2010fundamentals,Vierendeels2012}. Partitioned methods are inherently modular, and thus enable leveraging the extensive capabilities of contemporary CFD (Computational Fluid Dynamics) and CSM (Computational Structural Mechanics) solvers, e.g. turbulence modeling, advanced geometric modeling and meshing capabilities, contact-treatment, non-linear material modeling, High-Performance Computing (HPC) capabilities on diverse hardware configurations, etc. Partitioned methods therefore provide much greater flexibility in addressing diverse FSI problems compared to the monolithic approaches, and are better aligned with industrial workflows. Partitioned methods can generally be classified as loosely-coupled (or {\it staggered\/}) and strongly-coupled methods. In loosely coupled methods, the fluid and structure systems are solved only once per time step, while in strongly coupled methods, the fluid and solid subsystems are solved alternatingly until a prescribed convergence tolerance is reached. Such repeated, alternating solution of the fluid and solid subsystems is commonly referred to as {\it subiteration\/}. Loosely coupled methods are generally appropriate for weakly coupled FSI problems, such as those encountered in aeroelasticity~\cite{farhat2004cfd}, while strongly coupled methods are required for FSI problems that exhibit strong interaction, such as those encountered in cardiovascular applications. 

In the standard subiteration method, the fluid receives the velocity and displacement information from the structure in accordance with the kinematic interface condition, which can be conceived of as a Dirichlet-type boundary condition on the fluid subsystem, while the fluid in turn transfers traction to the wetted boundary of the solid in compliance with the dynamic interface condition, which can be regarded as a Neumann-type boundary condition on the solid subsystem~\cite{van2010fundamentals,Fernandez-moubachir}. Such a partition is called {\it Dirichlet--Neumann\/} ({\it {DN}}) coupling. Dirichlet--Neumann coupling is standard in partitioned FSI methods, carrying the advantage that it translates into canonical boundary conditions for the fluid and solid subsystems. The stability of the {DN} scheme is however conditional. The effect of the fluid on the solid can be characterized as an added mass to the solid~\cite{Stokes:1851qf}. In the {DN} partitioned procedure, this added mass is treated explicitly and, hence, if the added mass is relatively large compared to the actual structural mass, then the subiteration process exhibits prohibitively slow convergence, or it is unstable~\cite{van2009added,causin2005added}. The added-mass effect manifests itself in both compressible and incompressible fluids, but its character is fundamentally different for each: In compressible fluids, the added mass is proportional to the time-step size, while in incompressible fluids it is time-step independent and tends to a constant value in the limit of vanishing time-step 
size~\cite{van2009added,van2011partitioned}. If the added-mass effect is excessive, the subiteration process requires auxiliary stabilization techniques to effectuate or accelerate convergence, e.g. by relaxing the force or displacement data that is transferred, using a constant or dynamic~\cite{kuttler2} relaxation factor; by introducing {\it interface artificial compressibility} (IAC) of  the fluid in the vicinity of the interace~\cite{degrooteAC,raback2001fluid}; similarly, 
by increasing the diagonal dominance of the matrix representation of the continuity equation by incorporating a user-defined constant into the continuity equation~\cite{chimakurthi2018ansys,ansys2021}; or by a quasi-Newton 
procedure~\cite{Haelterman2009,Degroote2010,HAELTERMAN20169} or subspace-acceleration~\cite{Michler2005} procedure.

Another essential limitation of the {DN} scheme, emerges in FSI problems with an incompressible fluid that is subjected to a Dirichlet boundary condition on the velocity on its entire boundary, excluding the solid-fluid interface. The {DN} partition then results in a fluid subproblem that is constrained by Dirichlet conditions on its entire boundary, which engenders a Fredholm alternative~\cite{timo_van2015}: if the boundary data is compatible with the incompressibility constraint, the fluid subsystem possesses a solution that is unique up to a constant in the pressure; otherwise, a solution does not exist. A fundamental complication in {DN} schemes is that the structure subsystem is ignorant to the compatibility condition associated with the fluid subsystem and, hence, generally provides displacements that are incompatible, leading to non-existence of a fluid solution. This deficiency of {DN} schemes is commonly referred to as the 
{\it incompressibility dilemma}~\cite{Kuttler}. Several measures have been proposed to resolve the incompressibility dilemma: K\"uttler et al.~\cite{Kuttler} proposed to augment the structure system with an auxiliary constraint to impose the compatibility condition on the structural deformation. This constraint is imposed by means of a Lagrange multiplier, which can be interpreted as the excess pressure that is required to comply with the fluid-incompressibility constraint~\cite{timo_van2015,timo_thesis}. Bogaers et al.~\cite{bogaers2016evaluation} proposed to bypass the incompressibility dilemma by introducing artificial compressibility into the fluid subsystem. It is to be noted, however, that the results obtained by this procedure appear to deviate from incompressible benchmark results~\cite{Kuttler}. Another means of avoiding the compatibility condition associated with the {DN} scheme, is to modify the partitioning of the interface conditions, by imposing a Robin condition on the fluid subsystem at the fluid-solid interface, instead of a Dirichlet condition~\cite{Badia:2008fk,gerardo2010analysis,Astorino:2009eq,fernandez2013explicit}. The resulting coupling strategy is referred as a {\it Robin--Neumann\/} ({\it {RN}\/}) scheme. The Robin condition constitutes a natural boundary condition on the fluid subsystem, which implies that the deviation between the fluid velocity and the solid velocity is proportional to the deviation between the fluid traction and the solid traction. Upon convergence of the subiteration procedure for the {RN} scheme, the deviation between the fluid and solid traction vanishes and, hence, so does the deviation between the fluid and solid velocities at the interface. With an appropriate scaling of the relaxation parameter (i.e., the flow resistance) in the Robin condition, the Robin--Neumann scheme can be used to mitigate the added-mass effect. Moreover, by virtue of the fact that in the {RN} scheme the fluid subsystem is subject to a natural boundary condition on part of its boundary, {RN} coupling also overcomes the incompressibility dilemma. However, RN coupling schemes are generally intrusive with respect to existing commercial and open-source software packages, in terms of the data that is required in the RN boundary condition in the fluid subsystem.

\begin{figure}
\begin{center}
\includegraphics[width=0.6\textwidth]{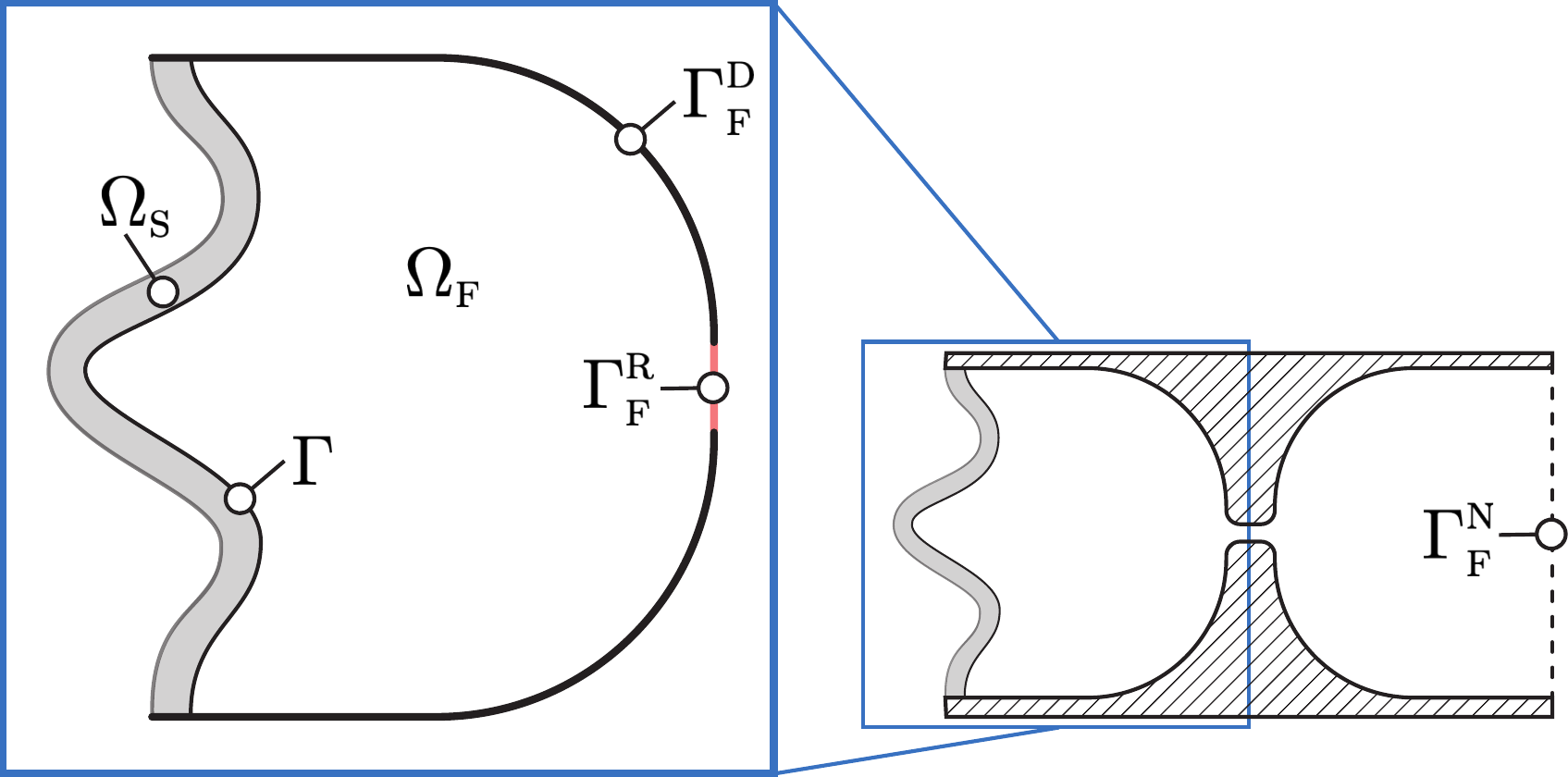}
\caption{Illustration of a fluid-structure-interaction problem with a nearly-closed fluid domain. If the gap between the left and right parts of the fluid domain is narrow, the flow resistance across the 
boundary~$\Gamma_{\FF}^{\textsc{r}}$ is large, left part of the fluid domain is nearly closed.
\label{fig:illustration}}
\end{center}
\end{figure}

The binary situation pertaining to the incompressibility dilemma in {DN} subiteration schemes is well understood. However, contemporary understanding of convergence issues that occur in FSI problems with nearly-closed incompressible fluid configurations 
is still incomplete. Even the designation {\it nearly-closed\/} (or {\it quasi-closed\/}~\cite{Kuttler}) is generally ambiguous. It has been conjectured~\cite{Kuttler} that a weakened version of the incompressibility dilemma applies in quasi-closed scenarios. However, the dichotomy implied by the Fredholm alternative is exclusive to fully closed configurations. It appears that an appropriate designation of a nearly-closed configuration of a fluid domain in an FSI problem, is a (part of) fluid domain that is enclosed by a flexible structure, a part of the boundary that carries Dirichlet boundary conditions on velocity (typically, a rigid wall), and a part that carries an arbitrarily large flow resistance; see the illustration in Figure~\ref{fig:illustration}. Such nearly-closed configurations occur in a wide variety of FSI problems. A typical example is provided by valves~\cite{Bavo2016,bogaers2016evaluation,Aissa-Berraies:2025fv}, where a flexible membrane connects an upstream and a downstream region. In the closed configuration of the valve, the upstream region is disconnected from the downstream region, and is fully closed. However, in nearly closed configurations, the upstream region is connected to the downstream region by an arbitrarily narrow gap, introducing an arbitrarily large resistance between the upstream and downstream regions. In a valve-closing scenario, instability of the {DN} subiteration procedure is to be  anticipated for the closed configuration, on account of the incompressibility dilemma. However, in practice one observes that the subiteration procedure fails well before the closed configuration is reached. It has been conjectured that pressure-correction schemes in the fluid solver, such as SIMPLE or PISO, can serve to mitigate convergence issues related to closed or nearly-closed configurations~\cite{bogaers2016evaluation}. However, if so, accuracy of the 
approximation and convergence of the {DN} subiteration procedure are vulnerable to details of the solver settings.

The objective of the present paper is to elucidate the convergence problems that occur for Dirichlet--Neumann subiteration schemes for FSI problems with nearly-closed configurations of the fluid domain and distinguish between the incompressibility dilemma pertaining to fluids with pure Dirichlet boundary conditions and the numerical difficulties corresponding to FSI problems with nearly-closed fluid subdomains. Based on a prototypical model problem, we establish that the near-closedness manifests as an {\it artificial added damping\/} in the structure subsystem. The effect of this artificial added damping on the convergence of the subiteration method with DN coupling can be represented by recursion of a Volterra operator. This Volterra operator is nonnormal, which results in distinct transient and asymptotic convergence behavior of the DN subiteration process and, potentially, in non-monotonous convergence~\cite{brummelen-borst,Trefethen1997}. In addition, we establish that the norm of the Volterra operator is proportional to the time step in the numerical procedure, which implies that the artificial added damping effect can be controlled by means of the time step.

The remainder of this paper is organized as follows. Section~\ref{sec:ProbFrom} presents the formulation of the considered class of nearly closed incompressible FSI problems. Section~\ref{sec:DNpartitioned} describes the {\it Dirichlet--Neumann\/} partitioned solution strategy and elaborates on the complications that emerge for closed and nearly-closed fluid domains. Based on a prototypical model problem, Section~\ref{sec:analytical} discusses the relation between flow resistance in nearly-closed FSI problems and the robustness and convergence of subiteration with DN coupling. Section~\ref{sec:num-ex} presents numerical experiments to examine the generalization of the  findings in the context of the model problem to more complex nearly-closed FSI problems. Finally, Section~\ref{sec:conclusion} presents concluding remarks.

\section{Problem formulation}
\label{sec:ProbFrom}
This section presents the formulation of the fluid-structure-interaction problem based on the traditional three-field formulation~\cite{farhat2004cfd}. The three-field formulation of FSI problems is standard, and the presentation in this section serves for coherence of the presentation, and to provide a setting to elucidate the notion of nearly-closed fluid domains in FSI.

We consider a fluid contiguous to a deformable solid. The FSI problem is set on a time interval $(0,t_{\fin})$ and a spatial domain~$\Omega\subset\IR^d$ ($d=2,3$). The domain~$\Omega$ consists of two complementary time-dependent subdomains~$\Omega_{\FF}(t)$ and~$\Omega_{\SS}(t)$, accommodating the fluid and solid, respectively. The fluid-solid interface~$\Gamma(t)=\partial\Omega_{\FF}(t)\cap\partial\Omega_{\SS}(t)$ corresponds to the intersection between the boundaries of the fluid and solid domains. We assume that the fluid and solid domains can be viewed as a time-dependent continuous deformation~$\bs{d}:(0,t_{\fin})\times\Omega\to\Omega$ acting on a reference configuration corresponding to the initial configuration, i.e. $\Omega_{\FF}(t)=\bs{d}(t,\hat{\Omega}_{\FF})$ 
and~$\Omega_{\SS}(t)=\bs{d}(t,\hat{\Omega}_{\SS})$ 
with $\hat{\Omega}_{\FF}=\Omega_{\FF}(0)$ and~$\hat{\Omega}_{\SS}=\Omega_{\SS}(0)$. We denote 
by~$\bs{d}_{\FF}=\bs{d}|_{{\hat{\Omega}}_{\FF}}$ and~$\bs{d}_{\SS}=\bs{d}|_{{\hat{\Omega}}_{\SS}}$ the deformation of the fluid and solid subdomains, respectively. Extending this notation,
in the remainder of this paper, quantities associated with the fluid (resp. solid) are generally indicated by a subscript ${\FF}$ (resp. ${\SS}$).

\subsection{Fluid subsystem} 
\label{sec:fluid}
We assume that the considered fluid is incompressible. The fluid flow is then described by 
a velocity field $\textbf{u}_{\FF}: (0,t_{\fin})\times\Omega_{\FF}(t)\to\IR^d$ and pressure field 
$p_{\FF}: (0,t_{\fin})\times\Omega_{\FF}(t)\to\IR^d$. To disambiguate the previous notation related to time-dependent functions on time-dependent domains, we note that such functions are unambiguously defined by their pullback to the reference configuration by the deformation~$\bs{d}_{\FF}(t,\cdot)$.
The velocity and pressure are subject to the incompressible Navier--Stokes equations:

\begin{subequations}
\label{eq:NS}
\begin{alignat}{3}
\rho_{\FF}\partial_t \boldsymbol{u}_{\FF} + 
\rho_{\FF} (\boldsymbol{u}_{\FF}\cdot  \nabla) \boldsymbol{u}_{\FF}+\nabla{}p_{\FF}
-\nabla \cdot \boldsymbol{\tau}_{\FF}&= 0 &\quad& \text{in }(0,t_{\fin})\times\Omega_{\FF}(t),
\label{eq:NS_mom}
\\
\nabla \cdot \boldsymbol{u}_{\FF}&=0 &\quad& \text{in }(0,t_{\fin})\times\Omega_{\FF}(t),
\label{eq:NS_cont}
\end{alignat}
\end{subequations}
where $\rho_{\FF}$ represents the fluid density and $\boldsymbol{\tau}_{\FF}$ denotes the viscous stress tensor according to:
\begin{equation}
\boldsymbol{\tau}_{\FF} =2\mu_{\FF}\bs{\varepsilon}(\bs{u}_{\FF})=\mu_{\FF} \big(\nabla \boldsymbol{u}_{\FF} + (\nabla \boldsymbol{u}_{\FF})^T\big),
\end{equation}
with $\boldsymbol{\varepsilon}(\bs{u})=\frac{1}{2}(\nabla \boldsymbol{u} + (\nabla \boldsymbol{u})^T)$ as the strain-rate tensor and $\mu_{\FF}$ as the dynamic viscosity. To account for the motion of the fluid domain, the Navier--Stokes equations~\eqref{eq:NS} are generally reformulated in the arbitrary Lagrangian-Eulerian (ALE) form on the reference configuration~\cite{van2010fundamentals, Donea1982,farhat2004cfd,Bazilevs:2008kk}. This reformulation is standard, and will not be further elaborated here.

Equations~\eqref{eq:NS} must be supplemented with suitable initial and boundary conditions. The interface conditions are discussed in Section~\ref{sec:intcon}. For the boundary conditions on the complementary part $\partial\Omega_{\FF}\setminus\Gamma$, we restrict our considerations 
to Dirichlet-, Neumann- and Robin-type boundary conditions, according to
\begin{subequations}
\label{eq:bcF}
\begin{alignat}{3}
\boldsymbol{u}_{\FF} & = \boldsymbol{u}^{\textsc{d}}_{\FF}
&\quad& \text{on }(0,t_{\fin})\times\Gamma^{\textsc{d}}_{\FF},
\label{eq:bcFD}
\\
p_{\FF}\bs{n}_{\FF}-\boldsymbol{\bs{\tau}_{\FF}}\boldsymbol{n}_{\FF} & =  \boldsymbol{t}_{\FF}^{\textsc{n}} 
&\quad& \text{on }(0,t_{\fin})\times\Gamma^{\textsc{n}}_{\FF},
\label{eq:bcFN}
\\
p_{\FF}\bs{n}_{\FF}-\boldsymbol{\bs{\tau}_{\FF}}\boldsymbol{n}_{\FF} 
-
\kappa_{\textsc{f}}\bs{u}_{\FF}
& =
\boldsymbol{t}_{\FF}^{\textsc{r}} 
&\quad& \text{on }(0,t_{\fin})\times\Gamma^{\textsc{r}}_{\FF},
\label{eq:bcFR}
\end{alignat}
\end{subequations}
respectively, with $\boldsymbol{n}_{\FF}$  as the exterior unit normal vector to the fluid domain, and $\boldsymbol{u}^{\textsc{d}}_{\FF}: (0,t_{\fin})\times\Gamma^{\textsc{d}}_{\FF}\to\IR^d $ and
$\boldsymbol{t}^{\textsc{n}}_{\FF}: (0,t_{\fin})\times\Gamma^{\textsc{n}}_{\FF}\to\IR^d$
and
$\boldsymbol{t}^{\textsc{r}}_{\FF}: (0,t_{\fin})\times\Gamma^{\textsc{r}}_{\FF}\to\IR^d$
prescribed velocity and traction data, respectively. The boundary sets
$\Gamma^{\textsc{d}}_{\FF}$, $\Gamma^{\textsc{n}}_{\FF}$ and $\Gamma^{\textsc{r}}_{\FF}$ 
are non-overlapping, and together with the interface $\Gamma(t)$ they provide
a covering of the entire fluid-domain boundary. The constant~$\kappa_{\textsc{f}}$ in the Robin condition~\eqref{eq:bcFR} is non-negative. Equation~\eqref{eq:bcFR} can be rearranged as
$\bs{u}_{\FF}
=\kappa_{\textsc{f}}^{-1}(p_{\FF}\bs{n}_{\FF}-\boldsymbol{\bs{\tau}_{\FF}}\boldsymbol{n}_{\FF}-\boldsymbol{t}_{\FF}^{\textsc{r}})$, conveying that the Robin condition imposes that the fluid velocity~$\bs{u}_{\FF}$ is
proportional to the deviation between the fluid traction~$p_{\FF}\bs{n}_{\FF}-\boldsymbol{\bs{\tau}_{\FF}}\boldsymbol{n}_{\FF}$ and the exogenous traction~$\boldsymbol{t}_{\FF}^{\textsc{r}}$. The proportionality factor~$\kappa_{\textsc{f}}$ (resp. its inverse $\kappa_{\textsc{f}}^{-1}$) represents flow resistance (resp. conductivity).

A suitable initial condition for the fluid subsystem consists of a specification of the initial velocity:
\begin{equation}
\label{eq:uF0}
\bs{u}_{\FF}(t=0)=\bs{u}_{\FF}^0\quad\text{in }\hat{\Omega}_{\FF}
\end{equation}
for suitable initial data~$\bs{u}_{\FF}^0:\hat{\Omega}_{\FF}\to\IR^d$.

\subsection{Structure subsystem}
\label{sec:structure}
We consider a large-displacement, i.e. geometrically nonlinear, formulation of the structure. The equation of motion for the structure then imposes:
\begin{equation}
\label{eq:solid}
   \rho_{\SS} \partial^2_t \boldsymbol{d}_{\SS}-  \nabla_{\bs{X}} \cdot \boldsymbol{P}_{\SS} = 0 \quad \text{in }(0,t_{\fin})\times\hat{\Omega}_{\SS}.
\end{equation}
where $\rho_{\SS}$ denotes the solid mass density in the reference configurations, $\nabla_{\bs{X}}$ represents the gradient operator acting in the reference configuration and $\nabla_{\bs{X}}\cdot(\cdot)$ the corresponding divergence operator, and $\bs{P}$ stands for the first Piola--Kirchhoff stress tensor. 
Equation~\eqref{eq:solid} must be furnished with a constitutive relation that relates the stress~$\boldsymbol{P}_{\SS}$ to the strain corresponding to~$\boldsymbol{d}_{\SS}$. We restrict our considerations to hyperelastic materials. Denoting by 
$\bs{F}_{\SS}=\nabla_{\bs{X}}\bs{d}_{\SS}$ the deformation gradient, and by $\bs{E}_{\SS}=\tfrac{1}{2}(\bs{F}^T_{\SS}\bs{F}_{\SS}-\bs{I})$ the Green--Lagrange strain tensor, a hyperelastic material is characterized by a stored-energy-density 
function~$W:=W(\bs{E}_{\SS})$, from which the second Piola--Kirchoff stress tensor is derived as $\bs{S}_{\SS}=\mathrm{d}W(\bs{E}_{\SS})/\mathrm{d}\bs{E}_{\SS}$, and the first Piola--Kirchhoff stress tensor as $\bs{P}_{\SS}=\bs{F}_{\SS}\bs{S}_{\SS}$. 

The interface conditions on the structure subsystem at the wetted boundary boundary~$\hat{\Gamma}\subset\partial\hat{\Omega}_{\SS}$ are elaborated in Section~\ref{sec:intcon}. We assume that on the complementary part, the structure is subjected to Dirichlet- and Neumann-type conditions:
\begin{subequations}
\label{eq:bc-solid}
\begin{alignat}{3}
\boldsymbol{d}_{\SS} =& \boldsymbol{d}_{\SS}^{\textsc{d}} 
&\quad& \text{on }(0,t_{\fin})\times\hat{\Gamma}{}^{\textsc{d}}_{\SS}
\label{eq:bc-solidD}
\\
\boldsymbol{P}_{\SS}\boldsymbol{N}_{\SS} =& \boldsymbol{T}{}^{\textsc{n}}_{\SS}  \
&\quad& \text{on }(0,t_{\fin})\times\hat{\Gamma}{}^{\textsc{n}}_{\SS}
\label{eq:bc-solidN}
\end{alignat}    
\end{subequations}
where $\boldsymbol{d}_{\SS}^{\textsc{d}}$ and~$\boldsymbol{T}{}^{\textsc{n}}_{\SS}$ represent prescribed deformation and traction data, respectively. Note that the latter is specified in the reference configuration. 

Suitable initial conditions for~\eqref{eq:solid} are provided by a specification of the initial deformation and initial velocity. Recalling that the reference configuration coincides with the initial configuration, and assuming that the structure is initially stationary, it holds that:
\begin{subequations}
\label{eq:ic-solid}
\begin{alignat}{3}
\boldsymbol{d}_{\SS}(t=0,\cdot) =& (\cdot)&\quad& \text{in }\hat{\Omega}_{\SS},
\\
\partial_t\boldsymbol{d}_{\SS}(t=0,\cdot) =& 0&\quad& \text{in }\hat{\Omega}_{\SS},
\end{alignat}
\end{subequations}
We insist that the boundary data in~\eqref{eq:bc-solidD} is compatible with the initial conditions, in the sense that $\boldsymbol{d}_{\SS}^{\textsc{d}}(t=0,\cdot)=(\cdot)$ and $\partial_t\boldsymbol{d}_{\SS}^{\textsc{d}}(t=0,\cdot)=0$.

\subsection{Fluid domain deformation}
\label{sec:moving_domain}
To accommodate the deformation of the structure domain, the fluid domain has to deform accordingly. 
Several approaches have been proposed to extend the deformation of the structural domain into the fluid domain. Typically, an auxiliary boundary-value problem must be solved on the fluid domain, e.g. a diffusion problem or an elastic-solid problem. In the latter case, the solid problem has no physical significance, and it is therefore generally referred to as a {\it pseudo-solid\/}. Corresponding formulations of FSI problems are referred to as {\it three-field formulations\/}~\cite{farhat2004cfd}, in view of the incorporation of a third field, viz. the deformation of the fluid domain, in addition to the solid subsystem and the fluid subsystem. 

Recalling the notation for the deformation of the fluid domain relative to the reference configuration, $\bs{d}_{\FF}$, the auxiliary diffusion problem writes:
\begin{subequations}
\label{eq:fludomap}
\begin{alignat}{3}
\nabla_{\boldsymbol{X}}\cdot(\gamma_{\FF}\bs{d}_{\FF})&=0&\quad&\text{in }(0,t_{\fin})\times\hat{\Omega}_{\FF},
\label{eq:meshdiff}
\\
\bs{d}_{\FF}(t,\cdot)&=(\cdot)&\quad&\text{on }(0,t_{\fin})\times(\partial\hat{\Omega}_{\FF}\setminus\hat{\Gamma}).
\label{eq:ds_bcD}
\end{alignat}    
\end{subequations}
For coherence of the presentation, the boundary condition associated with~\eqref{eq:meshdiff} at the fluid-structure interface is presented in Section~\ref{sec:intcon}.
Boundary condition~\eqref{eq:ds_bcD} implies that the complementary part of the fluid-domain boundary remains undeformed. This condition can be modified to include imposed deformations of the fluid domain, but we will not consider that further here. The diffusion 
coefficient~$\gamma_{\FF}$ can be used to localize the deformation near the fluid-structure 
interface~ \cite{lohner1996improved}. Alternatively, \eqref{eq:meshdiff} can be replaced by a pseudo-solid deformation analogous to~\eqref{eq:solid}. In numerical procedures, the stress-strain relation pertaining to the pseudo-solid can be used to control the mesh deformation corresponding to the domain motion, e.g. by Jacobian-based stiffening~\cite{Stein:2003no} to reduce the deformation of small elements in the vicinity of the structure.

\subsection{Interface conditions}
\label{sec:intcon}
The fluid, structure and fluid-domain-deformation subsystems are coupled at their mutual interface by kinematic and dynamic interface conditions. The kinematic condition comprises two parts, and imposes that the fluid and solid domains remain connected at the interface, and that the fluid velocity coincides with the structural velocity at the interface. The dynamic condition encodes the balance of the tractions exerted by the fluid and the structure on the interface.

Connectedness of the fluid and structure domain is accounted for by imposing that the deformation on the aggregate FSI domain, $\bs{d}:(0,t_{\fin})\times\Omega\to\Omega$, is continuous across the fluid solid interface and, hence, 
\begin{equation}
\label{eq:kin1}
\bs{d}_{\FF}|_{\hat{\Gamma}}=\bs{d}_{\SS}|_{\hat{\Gamma}}
\quad\text{on }(0,t_{\fin})\times\hat{\Gamma}.
\end{equation}
Condition~\eqref{eq:kin1} can be conceived of as an essential boundary condition for the fluid-domain-deformation equation~\eqref{eq:meshdiff}.

The second kinematic condition, imposing that the fluid and structure velocities coincide at the interface, yields
\begin{equation}
\label{eq:diric}
(\boldsymbol{u}_{\FF}\circ\,\bs{d}_{\FF})\big|_{\hat{\Gamma}}
= 
\partial_t \bs{d}_{\SS}\big|_{\hat{\Gamma}}
\quad\text{on }(0,t_{\fin})\times\hat{\Gamma}.
\end{equation}
The composition with the deformation~$\bs{d}_{\FF}$ in the left-hand side is necessary because the fluid velocity~$\bs{u}_{\FF}$ is defined in the current configuration, while the solid deformation~$\bs{d}_{\SS}$ that appears in the right-hand side of~\eqref{eq:diric} is defined in the reference configuration.

The traction exerted by the fluid on the interface corresponds to $p_{\FF}\bs{n}_{\FF}-\bs{\tau}_{\FF}\bs{n}_{\FF}$; cf.~\eqref{eq:bcFN}. Denoting by $\bs{\sigma}_{\SS}$ the Cauchy stress in the structure, and by $\bs{n}_{\SS}$ the external unit normal vector to the structure in the deformed configuration, the structure exerts traction $-(\bs{\sigma}_{\SS}\bs{n}_{\SS})\,\circ\,\bs{d}_{\SS}^{-1}$ on the interface in the current configuration. The dynamic condition insist that the following traction balance holds:
\begin{equation}
\label{eq:dyncon1}
p_{\FF}\bs{n}_{\FF}-\bs{\tau}_{\FF}\bs{n}_{\FF}-(\bs{\sigma}_{\SS}\bs{n}_{\SS})\,\circ\,\bs{d}_{\SS}^{-1}=0
\qquad\text{on }(0,t_{\fin})\times\Gamma(t)
\end{equation}
Recalling that the Cauchy stress and the first Piola-Kirchhoff stress are related via the deformation tensor by $\bs{P}_{\SS}=\det(\bs{F}_{\SS})\bs{\sigma}_{\SS}\bs{F}_{\SS}^{-T}$, with $\bs{F}_{\SS}^T$ the transpose of~$\bs{F}_{\SS}$ and~$\bs{F}_{\SS}^{-T}$ its inverse, it follows from~\eqref{eq:dyncon1} and Nanson's formula that for all suitable functions $\bs{w}:\Gamma(t)\to\IR^d$:
\begin{multline}
\label{eq:dyncon2}
\int_{\Gamma(t)}\bs{w}\cdot(p\bs{n}_{\FF}-\bs{\tau}_{\FF}\bs{n}_{\FF})\,\mathrm{d}\Gamma
=
\int_{\Gamma(t)}\bs{w}\cdot(\bs{\sigma}_{\SS}\bs{n}_{\SS})\,\circ\,\bs{d}_{\SS}^{-1}\,\mathrm{d}\Gamma
\\
=
\int_{\hat{\Gamma}}(\bs{w}\,\circ\,\bs{d}_{\SS})\cdot\bs{\sigma}_{\SS}\bs{F}^{-T}\bs{N}_{\SS}\,\det(\bs{F})\,\mathrm{d}\hat{\Gamma}
=
\int_{\hat{\Gamma}}(\bs{w}\,\circ\,\bs{d}_{\SS})\cdot\bs{P}_{\SS}\bs{N}_{\SS}\,\mathrm{d}\hat{\Gamma}
\end{multline}
with $\bs{n}_{\SS}$ the external unit normal vector in the current configuration. Equation~\eqref{eq:dyncon2} implies that the dynamic condition can be imposed by placing the left-hand side of~\eqref{eq:dyncon2} in the right-hand side of the weak formulation of the solid subsystem. More precisely, if the fluid traction is evaluated in a variationally consistent manner, the dynamic condition is enforced by continuity of the test function for the balance equations for linear momentum of the fluid and the structure; see, e.g.,~\cite{Brummelen:2012fk,van2010fundamentals,Brummelen:2021wt}.

\subsection{Nearly-closed fluid domains}
\label{sec:nearlyclosed}
In this paper we focus on FSI problems with an incompressible fluid in a nearly-closed domain. FSI problems of this type occur if (part of) the fluid domain is fully enclosed by an impermeable elastic structure, by a part carrying Dirichlet conditions according to~\eqref{eq:bcFD}, e.g. corresponding to a rigid wall or an imposed inflow velocity, and by a permeable part characterized by a large flow resistance. The permeable part of the boundary can be represented by a Robin condition~\eqref{eq:bcFR}, and the nearly-closed scenario corresponds to large values of the flow resistance~$\kappa_{\textsc{f}}$.

Figure~\ref{fig:illustration} presents an illustration of a nearly-closed FSI problem.
The fluid domain comprises two parts which are connected by a narrow channel. The left part of the fluid domain is enclosed by the interface~$\Gamma$ with the flexible structure in~$\Omega_{\SS}$, a rigid impermeable wall~$\Gamma_{\FF}^{\textsc{d}}$ which carries a homogeneous Dirichlet condition on the fluid velocity, and the part $\Gamma_{\FF}^{\textsc{r}}$ corresponding to the aperture to the right part of the fluid domain. The boundary of the right subdomain comprises a part $\Gamma_{\FF}^{\textsc{n}}$ on which a Neumann condition~\eqref{eq:bcFN} holds.
The channel connecting the fluid subdomains constitutes a flow resistance, and the resistance of the channel increases as the cross-sectional area of the channel decreases. The collective effect of the channel and the right part of the fluid domain on the left part can be represented by a Robin-type boundary condition conforming to~\eqref{eq:bcFR}, acting on the part~$\Gamma^{\textsc{r}}_{\FF}$ of the boundary of the left fluid domain corresponding to the aperture of the channel; see Figure~\ref{fig:illustration}. The parameter~$\kappa_{\textsc{f}}$ in~\eqref{eq:bcFR} then represents the resistance of the channel. If the channel is narrow, the flow resistance~$\kappa_{\textsc{f}}$ is large, and the coupling between the motion of the flexible structure and the flow in the left part of the fluid domain engenders a fluid-structure-interaction problem with a nearly-closed fluid domain.

For large~$\kappa_{\textsc{f}}$, the Robin condition~\eqref{eq:bcFR} can formally be conceived of as a penalty formulation for the homogeneous Dirichlet condition~$\boldsymbol{u}_{\FF}=0$. Hence, in the limit~$\kappa_{\textsc{f}}\to\infty$, the nearly-closed fluid domain formally degenerates to a fully closed domain, and the FSI problem exhibits the incompressibility dilemma.

The aforementioned incompressible-fluid--structure--interaction problem with a nearly-closed fluid domain, in which the near-closedness of the fluid domain emerges from a large flow resistance in the connection to the open part of the fluid domain, is representative of FSI problems in a variety of engineering applications, notably, valve systems, e.g. mitral or aortic valves in cardiovascular mechanics~\cite{Bavo2016} or reed, ball or diaphragm valves in industrial applications~\cite{bogaers2016evaluation}. Valve systems correspond to 
fluid-structure-contact-interaction (FSCI) problems~\cite{Ager2019}, in which the contact leads to sealing of the valve, causing a separation of the fluid domain into two disconnected parts, viz. a closed upstream region and an open downstream region. In numerical procedures for FSCI problems, a finite gap in the contact region is usually maintained, to avoid a collapse of the fluid mesh. This finite gap then constitutes a large flow resistance. Instead of fully sealing the valve, some minute leakage is allowed~\cite{Vierendeels2012}. Alternatively, a porous-media model can be introduced in the contact region~\cite{Ager2019}, to further increase the flow resistance in the gap and accordingly reduce the leakage.

\section{The Dirichlet-Neumann partitioned coupling scheme for FSI}
\label{sec:DNpartitioned}
Numerical approximation of the aggregated FSI problem generally involves spatial and temporal discretization of the fluid, structure and fluid-domain-deformation subsystems, introduced in Secs.~\ref{sec:fluid}, \ref{sec:structure} and~\ref{sec:moving_domain}, respectively, subject to the interface conditions in~\ref{sec:intcon} and auxiliary initial and boundary conditions.
Spatio-temporal discretization leads to a sequence of non-linear algebraic systems, where each non-linear system in the sequence is associated with a time step. The composite configuration of the FSI problem carries over to the considered non-linear algebraic systems, which are of the form:
\begin{equation}
\label{eq:coupled1}
 \begin{aligned}
\tilde{\boldsymbol{R}}_{\SS}(\tilde{\boldsymbol{q}}{}^n_{\SS},\tilde{\boldsymbol{q}}{}^n_{\FF})&=0     
\\
\tilde{\boldsymbol{R}}_{\FF}(\tilde{\boldsymbol{q}}{}^n_{\SS},\tilde{\boldsymbol{q}}{}^n_{\FF},\tilde{\boldsymbol{q}}{}^n_{\textsc{m}})&=0     
\\
\tilde{\boldsymbol{R}}_{\textsc{m}}(\tilde{\boldsymbol{q}}{}^n_{\SS},\tilde{\boldsymbol{q}}{}^n_{\textsc{m}})&=0     
 \end{aligned}   
\end{equation}
where 
$\tilde{\boldsymbol{q}}{}^n_{\SS},\tilde{\boldsymbol{q}}{}^n_{\FF},\tilde{\boldsymbol{q}}{}^n_{\textsc{m}}$
represent the discrete variables associated with the structure, fluid and fluid-domain deformation, respectively, and 
$\tilde{\boldsymbol{R}}_{\SS},\tilde{\boldsymbol{R}}_{\FF},\tilde{\boldsymbol{R}}_{\textsc{m}}$ represent the residual equations for the various subsystems. It is to be noted that the fluid-domain deformation depends on the structure variables via the boundary condition~\eqref{eq:kin1}, but it does not depend on the fluid variables. Moreover, the structure subsystem is independent of the fluid-domain deformation. 

Depending on the setup of the subsystems, the fluid-domain deformation is generally regarded as part of, and merged with, either the fluid or structure subsystem. If the domain deformation corresponds to a pseudo solid which is discretized in the same manner as the structure subsystem, then the pseudo solid equations are naturally paired with the structure subsystem, while if dedicated domain-map techniques are provided by the fluid-domain solver, then the domain deformation is naturally merged with the fluid subsystem. It is to be noted that the latter is standard in commercial CFD solvers,
in view of the fact that the ALE formulation for flow problems on moving domains is intrinsically coupled to the deformation of the fluid domain.
In either case, the fluid-domain-deformation variables are integrated into the structure or fluid variables, and the domain-map equations are integrated into the structure or fluid subsystems, reducing the three-field coupled system in~\eqref{eq:coupled1} to:
\begin{equation}
\label{eq:coupled}
 \begin{aligned}
\boldsymbol{R}_{\SS}(\boldsymbol{q}^n_{\SS},\boldsymbol{q}^n_{\FF})&=0     
\\
\boldsymbol{R}_{\FF}
(\boldsymbol{q}^n_{\SS},\boldsymbol{q}^n_{\FF})&=0     
 \end{aligned}   
\end{equation}
The coupled system~\eqref{eq:coupled} is intrinsically modular: under the standard assumption that the discretized fluid and structure subsystems are well posed, for given $\boldsymbol{q}^n_{\FF}$ (resp. $\boldsymbol{q}^n_{\SS}$),
Equation~(\ref{eq:coupled}$_1$) (resp.~(\ref{eq:coupled}$_2$)) can be solved for 
$\boldsymbol{q}^n_{\SS}$ (resp. $\boldsymbol{q}^n_{\FF}$).

In this section, we consider solution procedures for the aggregated algebraic problem~\eqref{eq:coupled} and, in particular, the Dirichlet--Neumann partitioned solution procedure.

\subsection{Partitioned versus monolithic methods}
\label{sec:partmon}
To motivate the use of partitioned solution procedures for FSI, we briefly consider their relation to monolithic solution procedures. Numerical solution procedures for Fluid-Structure-Interaction problems can generally be classified into monolithic and partitioned methods. In monolithic approaches, the fluid and solid subsystems are solved simultaneously~\cite{Michler:2004vl,Heil2004}, i.e. the aggregated system~\eqref{eq:coupled} is solved by means of a solution procedure that disregards it modular character, e.g. a Newton procedure in which the linear tangent problems are solved by means of a direct solver. Monolithic methods offer, in principle, parameter-independent stability and convergence. However, the systems of linear-algebraic equations that need to be solved in monolithic procedures are generally large, non-sparse and severely ill-conditioned~\cite{Richter:2015kq}. A practically more important downside of monolithic methods is that these are inherently non-modular and, consequently, highly intrusive with respect to existing simulation tools for CFD and CSM. 

Partitioned methods~\cite{van2011partitioned,farhat2004cfd} retain and exploit the modularity of~\eqref{eq:coupled}, by solving~(\ref{eq:coupled}$_1$) and~(\ref{eq:coupled}$_2$) alternatingly until a prescribed convergence criterium is satisfied. The baseline partitioned procedure for~\eqref{eq:coupled}, referred to as {\it subiteration\/}, comprises the following iteration: for given suitable initial 
estimates~$\boldsymbol{q}^{n,0}_{\SS}$ and~$\boldsymbol{q}^{n,0}_{\FF}$ for the solution in time step~$n$, e.g. obtained from extrapolation of the solution in previous time steps, repeat
\begin{equation}
\label{eq:subiteration}
 \begin{aligned}
\boldsymbol{R}_{\SS}(\boldsymbol{q}^{n,k}_{\SS},\boldsymbol{q}^{n,k-1}_{\FF})&=0     
\\
\boldsymbol{R}_{\FF}
(\boldsymbol{q}^{n,k}_{\SS},\boldsymbol{q}^{n,k}_{\FF})&=0 
 \end{aligned}   
\end{equation}
for $k=1,2,\ldots$. It is important to note that 
$\boldsymbol{q}^{n,k}_{\SS}$ and~$\boldsymbol{q}^{n,k}_{\FF}$ can be obtained 
from~\eqref{eq:subiteration} by forward substitution. For this reason, the subiteration procedure is sometimes referred to as {\it Gauss--Seidel iteration\/}. The main advantage of partitioned procedures is that these effectively separate the FSI problem into subproblems associated with the structure and fluid subsystems, and data transfer between these subsystems.
By virtue of the fact that the structure and fluid subsystems are treated separately, separate solvers can be used for the structure and fluid and, hence, existing CSM and CFD simulation capabilities can be leveraged. The main disadvantage of partitioned methods is that their convergence behavior generally depends on problem parameters and, specifically, on the ratio of the added mass of the fluid to the mass of the structure~\cite{causin2005added,van2009added,Forster:2007aa}. The convergence behavior of subiteration generally deteriorates with increasing fluid-to-structure mass ratio, and if the mass ratio is too large, the iteration in~\eqref{eq:subiteration} will diverge. To enhance the robustness and convergence of partitioned methods, several approaches have been suggested~\cite{van2011partitioned}, such as basic under-relaxation, Aitken's method~\cite{Mok,Aitken:1926fk}, quasi-Newton methods~\cite{DEGROOTE2009793,HAELTERMAN20169}, and multi-grid~\cite{Richter:2015kq,Brummelen:2008an}.

\subsection{Dirichlet--Neumann coupling}
The partitioning of the aggregated FSI problem 
into subsystems associated with the fluid and structure in~\eqref{eq:coupled}, requires a partition and allocation of the interface conditions in Section~\ref{sec:intcon}. The standard partition consists in imposing the kinematic conditions~\eqref{eq:kin1} and~\eqref{eq:diric} as Dirichlet (essential) boundary conditions on the fluid-domain-deformation~\eqref{eq:fludomap} and 
fluid~\eqref{eq:NS} subsystems, respectively, and imposing the dynamic condition~\eqref{eq:dyncon2} as a Neumann (natural) boundary condition on the structure subsystem. In the subiteration procedure~\eqref{eq:subiteration}, this 
implies that the structure subsystem transfers its deformation data to the fluid and, in turn, the fluid transfers its traction data to the structure. The resulting coupling scheme is referred to as subiteration with 
{\it Dirichlet--Neumann (DN) coupling\/}. The advantage of the DN coupling scheme is that it translates into standard boundary-value problems for the fluid and structure subsystems.
Subiteration with DN coupling generally also forms the basic building block of more advanced partitioned iterative procedures.

\begin{remark}
\label{rem:Robin}
Alternate splittings of the interface conditions can be considered. For instance, the kinematic condition~\eqref{eq:diric} and dynamic condition~\eqref{eq:dyncon1} can be combined into:
\begin{equation}
\label{eq:FSI_Robin}
\big(\boldsymbol{u}_{\FF}
-
\partial_t\boldsymbol{d}_{\SS}\,\circ\,\boldsymbol{d}_{\SS}^{-1}\big)
{}
\big|_{\Gamma}
=
\kappa_{\textsc{f}}
\big((p_{\FF}\bs{n}_{\FF}-\bs{\tau}_{\FF}\bs{n}_{\FF})-(\bs{\sigma}_{\SS}\bs{n}_{\SS})\,\circ\,\bs{d}_{\FF}^{-1}\big)
\end{equation}
for some suitable coefficient $\kappa_{\textsc{f}}>0$. Equation~\eqref{eq:FSI_Robin} can be imposed as a Robin (mixed) boundary condition on the fluid subsystem at the interface. Conditition~\eqref{eq:FSI_Robin} and~\eqref{eq:dyncon1} are equivalent to~\eqref{eq:kin1} and~\eqref{eq:dyncon1}. The subiteration procedure with~\eqref{eq:FSI_Robin} imposed on the fluid subsystem and~\eqref{eq:dyncon1} (or, equivalently, \eqref{eq:dyncon2}) imposed on the structure subsystem is referred to as {\it Robin--Neumann (RN) coupling\/}~\cite{fernandez2013explicit,Badia:2008fk,Astorino:2009eq}. RN coupling has various favorable properties over DN coupling, but it is non-standard in FSI solution procedures, and generally it is intrusive with respect to existing commercial and open-source software.
\end{remark}

The subiteration scheme with DN coupling is summarized in Algorithm~\ref{alg:DN_subiteration}. In the algorithm, we assume that the fluid-domain deformation is incorporated in the fluid subsystem, which is the usual arrangement in most open-source and commercial software with FSI capabilities. Before entering the subiteration loop in line~8, the states of the fluid and structure subsystems are initialized to the solution of the previous time step, which serves as an initial estimate for the solution in the current time step. The transfer operations in lines 10 and~12 do not generally require transfer of the complete fluid and structure states, but only require transfer of the traction data at the fluid-structure interface from the fluid to the structure (line 10) and of the deformation and velocity data at the fluid-structure interface from the structure to the fluid (line 12). In line 11, the structure subsystem is solved for a new deformation field. Under the assumption that the fluid-domain deformation is integrated in the fluid subsystem, in line 13 the fluid subsystem is solved for the fluid-domain deformation and the velocity and pressure of the fluid. The optional acceleration/stabilization step in line 14 corresponds to an update of the form
\begin{equation}
(\boldsymbol{q}_{\SS}^{n},\boldsymbol{q}_{\FF}^{n})^k_{*}
=\mathcal{A}\big((\boldsymbol{q}_{\SS}^{n},\boldsymbol{q}_{\FF}^{n})^0,(\boldsymbol{q}_{\SS}^{n},\boldsymbol{q}_{\FF}^{n})^1,\ldots,(\boldsymbol{q}_{\SS}^{n},\boldsymbol{q}_{\FF}^{n})^{k-1},(\boldsymbol{q}_{\SS}^{n},\boldsymbol{q}_{\FF}^{n})^k\big)
\end{equation}
with $\mathcal{A}$ some suitable operator,
in which the most recent and all previous iterates are used to obtain an improved approximation, $(\boldsymbol{q}_{\SS}^{n},\boldsymbol{q}_{\FF}^{n})^k_{*}$. For instance, in the case of simple under relaxation, the operator~$\mathcal{A}$ corresponds to a convex combination of the two most recent iterates according to:
\begin{multline}
\mathcal{A}\big((\boldsymbol{q}_{\SS}^{n},\boldsymbol{q}_{\FF}^{n})^0,(\boldsymbol{q}_{\SS}^{n},\boldsymbol{q}_{\FF}^{n})^1,\ldots,(\boldsymbol{q}_{\SS}^{n},\boldsymbol{q}_{\FF}^{n})^{k-1},(\boldsymbol{q}_{\SS}^{n},\boldsymbol{q}_{\FF}^{n})^k\big)    
\\
=
\alpha(\boldsymbol{q}_{\SS}^{n},\boldsymbol{q}_{\FF}^{n})^k
+
(1-\alpha)
(\boldsymbol{q}_{\SS}^{n},\boldsymbol{q}_{\FF}^{n})^{k-1}
\end{multline}
with $0<\alpha\leq{}1$.
The acceleration/stabilization can also be applied to only the structure or fluid variables or, in fact, to one of the interface-data items that is transferred between the fluid and the structure; see, e.g., \cite{van2011partitioned,Mok,DEGROOTE2009793,HAELTERMAN20169}.
\begin{algorithm}
\caption{Subiteration with Dirichlet--Neumann coupling.\label{alg:DN_subiteration}}
\begin{algorithmic}[1]
\State \textbf{Input}: $\Delta{}t>0$  (time step), $t_{\textsc{fin}}>0$ (final time), $\boldsymbol{q}_{\SS}^0,\boldsymbol{q}_{\FF}^0$ (initial conditions for structure and fluid), $\text{tol}$ (tolerance)
\State \textbf{Initialize}: $n=0$ (time-step counter), $k=0$ (subiteration counter), $t_n=0$ (time level)
\State \textbf{While} $t_n < t_{\textsc{fin}}$
\State \quad $n = n+1$
\State \quad $t_n= t_n + \Delta t$
\State \quad $k=0$
\State \quad $\boldsymbol{q}_{\SS}^{n,0}=\boldsymbol{q}_{\SS}^{n-1},\boldsymbol{q}_{\FF}^{n,0}=\boldsymbol{q}_{\FF}^{n-1}$ (initialize states in current time step by solution from previous time step)
\State \quad \textbf{Repeat}
\State \quad \quad $k=k+1$
\State \quad \quad Transfer traction of fluid state $\boldsymbol{q}^{n,{k-1}}_{\FF}$ to structure solver
\State \quad \quad Solve $\boldsymbol{R}_{\SS}(\boldsymbol{q}^{n,k}_{\SS},\boldsymbol{q}^{n,k-1}_{\FF})=0$ for $\boldsymbol{q}^{n,k}_{\SS}$ (solve structure subsystem subject to dynamic condition~\eqref{eq:dyncon1})
\State \quad \quad
Transfer interface displacement and velocity of structure state $\boldsymbol{q}^{n,k}_{\SS}$ to fluid solver
\State \quad \quad Solve $\boldsymbol{R}_{\FF}(\boldsymbol{q}^{n,k}_{\FF},\boldsymbol{q}^{n,k}_{\SS})=0$ for $\boldsymbol{q}^{n,k}_{\FF}$ (solve fluid subsystem subject to kinematic conditions~\eqref{eq:kin1} and~\eqref{eq:diric})
\State \quad \quad Accelerate/stabilize iteration (optional)
\State \quad \textbf{Until} 
$\|(\boldsymbol{q}_{\SS}^{n},\boldsymbol{q}_{\FF}^{n})^k-(\boldsymbol{q}_{\SS}^{n},\boldsymbol{q}_{\FF}^{n})^{k-1}\|<\text{tol}$
\State \quad Set $(\boldsymbol{q}_{\SS}^{n},\boldsymbol{q}_{\FF}^{n})=
(\boldsymbol{q}_{\SS}^{n},\boldsymbol{q}_{\FF}^{n})^k$ (store solution for time step~$n$)
\State \textbf{End While}
\vspace{10pt}
\end{algorithmic}
\end{algorithm}

\subsection{DN coupling and (nearly-)closed fluid domains}
\label{sec:DN}
In the DN coupling scheme, the fluid subsystem is subjected to a Dirichlet boundary condition at the fluid-structure interface. An auxiliary compatibility condition then emerges if the fluid domain is {\it closed\/}, i.e. if the fluid is also subject to a Dirichlet boundary condition of the form~\eqref{eq:bcFD} on the complementary part of its boundary, 
$\Gamma_{\FF}^{\textsc{D}}=
\operatorname{int}(\partial\Omega_{\FF}\setminus\Gamma)$.
Such a situation occurs, for instance, if the fluid domain is enclosed by the fluid-structure interface, rigid walls, and an inlet with an imposed flow profile or, in the absence of rigid parts, if a flexible structure is inflated by an imposed inlet velocity~\cite{timo_van2015,Kuttler}. On account of the incompressiblity of the fluid and the kinematic condition~\eqref{eq:diric}, the following sequence of identities holds:
\begin{equation}
\label{eq:compatibility}
0=\int_{\Omega_{\FF}}\nabla\cdot\boldsymbol{u}_{\FF}
=
\int_{\partial\Omega_{\FF}}\boldsymbol{u}_{\FF}\cdot\boldsymbol{n}_{\FF}
=
\int_{\Gamma_{\FF}^\textsc{D}}
\boldsymbol{u}_{\FF}^\textsc{D}\cdot\boldsymbol{n}_{\FF}
+
\int_{\Gamma}\big((\partial_t\boldsymbol{d}_{\SS})\circ\boldsymbol{d}_{\SS}^{-1}\big)\cdot\boldsymbol{n}_{\FF}
\end{equation}
Equation~\eqref{eq:compatibility} conveys a constraint on the structure deformation at the fluid-structure interface in relation to the Dirichlet data on the complementary part of the fluid boundary. However, in the DN coupling scheme, the structure subsystem is ignorant of the data in the fluid subsystem, and hence the compatibility condition~\eqref{eq:compatibility} is generally violated. A consequence of this incompatibility of the structure deformation, is that the fluid subproblem does not admit a solution. Because this non-existence of a fluid solution is associated with the incompressibility of the fluid, it is generally referred to as the {\it incompressibility 
dilemma\/}~\cite{Kuttler}.

The compatibility condition associated with closed, incompressible flow problems of the above type, gives rise to a so-called Fredholm alternative: If~\eqref{eq:compatibility} is violated, the flow problem does not admit a solution. If~\eqref{eq:compatibility} is satisfied, a solution exists, but it is non-unique.
The non-uniqueness of the flow solution can be inferred from the fact that the pressure in~\eqref{eq:NS} does not appear in the boundary conditions of the problem and appears in~\eqref{eq:NS_mom} only under the gradient. Hence, the pressure is determined only up to a constant, i.e. if $(\boldsymbol{u},p)$ is a solution to the closed incompressible flow problem, then so is $(\boldsymbol{u},p+c)$ for arbitrary~$c\in\IR$;
see also~\cite{timo_van2015}.

The compatibility condition can be incorporated in the fluid and structure subsystems by means of Lagrange multipliers~\cite{timo_van2015}. On the fluid side, the Lagrange multiplier is associated with an auxiliary constraint that determines the pressure level. On the structure side, the Lagrange multiplier is associated with an auxiliary constraint that imposes the volumetric compatibility of the structure deformation with the fluid. It should be noted that the structure Lagrange multiplier can be interpreted as a uniform excess pressure on the wetted boundary that adjusts the deformation of the structure according to the compatibility condition~\cite{timo_van2015}. Despite the fact that the Langrange multipliers correspond to scalar constraints, their incorporation in existing software is generally severely intrusive. In addition, the constraint on the structure displacement requires transfer of non-standard data from the fluid to structure, which impairs the modularity of the partitioned approach. 

Alternate means of managing the compatibility condition are to introduce artificial compressibility in the fluid~\cite{Bogaers2015,degrooteAC,raback2001fluid}, or to replace the Dirichlet boundary condition on the fluid subsystem at the fluid-structure interface by a Robin condition; see Remark~\ref{rem:Robin}. It is to be noted, however, that if the compressibility in the artificial-compressibility method is reduced or the flow resistance in the Robin--Neumann coupling is increased, the pressure in the fluid becomes increasingly sensitive to volumetric deviations in the fluid domain and, hence, the stability of the partitioned iteration deteriorates. 

If the fluid domain is nearly closed, according to the description in Section~\ref{sec:nearlyclosed}, the dichotomy of the Fredholm alternative does not apply. Instead, the pressure in the fluid subsystem exhibits a strong sensitivity to volume-rate deviations in the fluid domain. To elucidate this dependence, we reconsider the identities in~\eqref{eq:compatibility}, but now assuming that part of the boundary of the fluid domain is furnished with a Robin boundary condition~\eqref{eq:bcFR}. It then follows that
\begin{multline}
\label{eq:compatibility2}
\int\nolimits_{{\Gamma}_{\FF}^{\textsc{r}}}
\big(p_{\FF}-\boldsymbol{n}_{\FF}\cdot\boldsymbol{\tau}_{\FF}\boldsymbol{n}_{\FF}
-
\boldsymbol{n}_{\FF}\cdot\boldsymbol{t}_{\FF}^{\textsc{r}}
\big)
=
\kappa_{\textsc{f}}
\int\nolimits_{{\Gamma}_{\FF}^{\textsc{r}}}\boldsymbol{u}_{\FF}\cdot\boldsymbol{n}_{\FF}
\\
=
-\kappa_{\textsc{f}}\bigg(
\int\nolimits_{\Gamma_{\FF}^\textsc{D}}
\boldsymbol{u}_{\FF}^\textsc{D}\cdot\boldsymbol{n}_{\FF}
+
\int\nolimits_{\Gamma}\big((\partial_t\boldsymbol{d}_{\SS})\circ\boldsymbol{d}_{\SS}^{-1}\big)\cdot\boldsymbol{n}_{\FF}
\bigg)
=-\kappa_{\textsc{f}}\mathcal{V}'
\end{multline}
where $\mathcal{V}'$ represents the volume-rate deviation, i.e. the difference between the inflow (resp. outflow) through the Dirichlet boundary per~\eqref{eq:bcFD} and the rate of expansion (resp. contraction) of the fluid domain due to the deformation at the fluid-structure interface corresponding to the kinematic condition~\eqref{eq:diric}. To facilitate the presentation, we regard a specific scenario in which the normal component of velocity,
$\boldsymbol{u}_{\FF}\cdot\boldsymbol{n}_{\FF}$,
and
pressure,~$p_{\FF}$, are uniform at~$\Gamma_{\FF}^{\textsc{r}}$, and the tangential component of the velocity, $\boldsymbol{n}_{\FF}\times\boldsymbol{u}_{\FF}\times\boldsymbol{n}_{\FF}$,  vanishes. Such a scenario can be constructed by a suitable arrangement of the boundary data $\boldsymbol{u}_{\FF}^{\textsc{d}}$ and $\boldsymbol{t}_{\FF}^{\textsc{r}}$. From the fact that~$p_{\FF}$ only appears under the gradient in~\eqref{eq:NS_mom}, and in~\eqref{eq:bcFR}, one can infer that if $(\boldsymbol{u}_{\FF},p_{\FF})$ solves~\eqref{eq:NS} subject to boundary conditions~\eqref{eq:bcFD} and~\eqref{eq:bcFR} and kinematic interface condition~\eqref{eq:diric} for some reference resistance coefficient~$\kappa_{\textsc{f}}^{\star}>0$, then $(\boldsymbol{u}_{\FF},p_{\FF}+\lambda)$ ($\lambda\in\IR$) solves the same boundary value problem for~$\kappa_{\textsc{f}}>0$ with, in particular, $\lambda=(\kappa_{\textsc{f}}-\kappa_{\textsc{f}}^{\star})\,\boldsymbol{u}_{\FF}\cdot\boldsymbol{n}_{\FF}$. Therefore, it holds that
\begin{equation}
\label{eq:voldev}
\lambda=
-\frac{\kappa_{\textsc{f}}-\kappa_{\textsc{f}}^{\star}}{\operatorname{meas}(\Gamma_{\FF}^{\textsc{r}})}\mathcal{V}'
\end{equation}
with $\operatorname{meas}(\Gamma_{\FF}^{\textsc{r}})$ the surface measure of the Robin boundary. Equation~\eqref{eq:voldev} conveys that for nearly-closed fluid domains, i.e. in the limit $\kappa_{\textsc{f}}\to\infty$, non-zero volume-rate deviations lead to unbounded variations in the pressure level,~$\lambda$. This strong sensitivity of the pressure level to volume-rate deviations leads to instability of the subiteration with DN coupling in the case of nearly-closed fluid domains.

\section{Model problem: the leaky piston}
\label{sec:analytical}
The inadequacy of subiteration with DN coupling for FSI problems with an incompressible fluid in a fully closed fluid domain, caused by the incompressibility dilemma, is well understood; see, e.g.,\cite{Kuttler,timo_thesis}. The exposition in Section~\ref{sec:DN} imparts that DN coupling is also unsuitable for FSI problems with an incompressible fluid in a nearly closed fluid domain, as described in Section~\ref{sec:nearlyclosed}, on account of severe sensivity of the fluid pressure to volume-rate deviations, and corresponding instability of the subiteration procedure.

In the present section, we elucidate the properties of subiteration with DN coupling for a simple model problem. The model problem pertains to a modification of the classical piston problem~\cite{Piperno1995THESIS}, in which the ideal gas in the setup in~\cite{Piperno1995THESIS} is replaced by an incompressible fluid, and the fixed lid of the cylinder in which the piston moves is permeable.

\subsection{Leaky piston problem}
\label{sec:leaky}
\begin{figure}
\begin{center}
\includegraphics[width=0.6\textwidth]{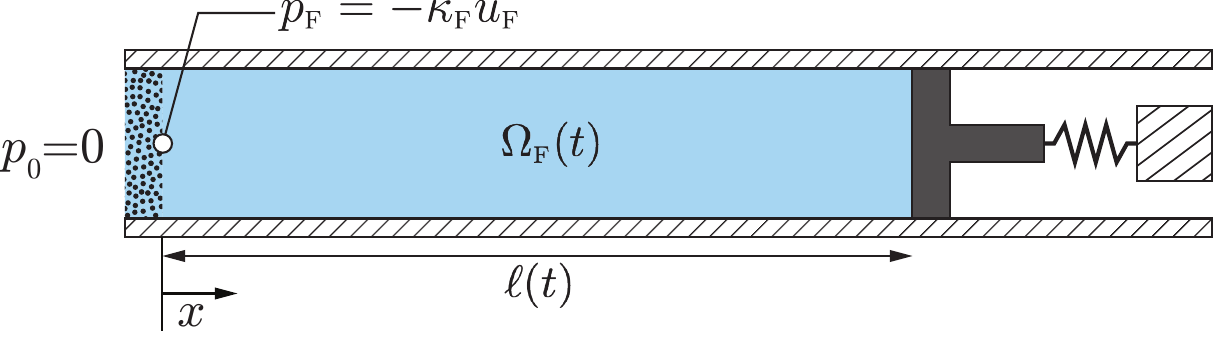}
\caption{Illustration of the {\it leaky-piston\/} model problem. The cylinder is at its left boundary connected to a reservoir at constant pressure via a permeable cover, causing a flow resistance described by a Robin-type condition. At the right boundary, the cylinder is closed by a piston which is connected to the environment by an elastic spring.\label{fig:LeakyPiston}}    
\end{center}
\end{figure}

We consider an incompressible, viscous fluid in a cylinder; see the illustration in Figure~\ref{fig:LeakyPiston}. At the left boundary, the cylinder is closed by a permeable cover. In view of the permeability of the cover, the cylinder is referred to as {\it leaky\/}. At the right boundary, the cylinder is closed by a freely moving piston. The piston is connected to an immobile fixture by means of a spring. Assuming that the fluid exhibits free slip along the lateral boundaries of the cylinder, 
and that the fluid velocity and pressure are uniform in the lateral directions,
the FSI problem corresponding to the fluid in the cylinder in connection with the piston admits a one-dimensional representation. In this one-dimensional representation, we denote by $\Omega_{\FF}(t)=(0,\ell(t))$ the domain occupied by the fluid. Denoting by $x$ the longitudinal coordinate and by $u_{\FF}:(0,t_{\textsc{fin}})\times\Omega_{\FF}(t)\to\IR$ the longitudinal velocity of the fluid, the incompressibility condition~\eqref{eq:NS_cont} reduces to $\partial_xu_{\FF}(t,x)=0$ and, hence, the velocity $u_{\FF}(t,\cdot)=:u_{\FF}(t)$ is uniform in the spatial dependence. The momentum equation for the fluid then reduces to
\begin{equation}
\label{eq:hyperb}
    \rho_{\FF}u'_{\FF} + \partial_x p_{\FF} =0
    \quad\text{in }(0,t_{\textsc{fin}})\times\Omega_{\FF}
\end{equation}
with $(\cdot)'$ as the derivative.
The permeable lid at the left-hand side is modelled by means of a homogeneous Robin-type boundary condition~\eqref{eq:bcFR}. Noting that the viscous part vanishes for spatially-uniform velocity, one can infer that $p_{\FF}(t,0)=-\kappa_{\textsc{f}}{}u_{\FF}(t)$. 
By integrating~\eqref{eq:hyperb} in the spatial dependence, one then obtains:
\begin{equation}
\label{eq:PS0}
\ell(t)\rho_{\FF}u_{\FF}'=p_{\FF}(t,\ell(t))+\kappa_{\textsc{f}}{}u_{\FF}(t)    
\quad
\text{in }(0,t_{\textsc{fin}})
\end{equation}

We next consider the piston, representing the structure subsystem. Denoting by $m_{\SS}$ and $\kappa_{\SS}$ the mass of the piston and the spring constant (both per unit area), and by $d_{\SS}:(0,t_{\textsc{fin}})\to\IR$ the displacement of the piston relative to its equilibrium position, the equation of motion for the piston is:
\begin{equation}
\label{eq:massspring}
m_{\SS}d_{\SS}''+\kappa_{\SS}d_{\SS} =p_{\FF}(t,\ell(t))
\quad\text{in }(0,t_{\textsc{fin}})
\end{equation}
 The right-hand side of~\eqref{eq:massspring} represents the load exerted by the fluid on the piston in accordance with the dynamic condition~\eqref{eq:dyncon1}. The ordinary differential equation~\eqref{eq:massspring} is furnished with initial conditions:
 \begin{equation}
 \label{eq:massspring_IC}
d_{\SS}(0)=\ell_0,
\qquad
d_{\SS}'(0)=u_0,
 \end{equation}
for certain initial deformation and velocity data, $\ell_0\in\IR$ and $u_0\in\IR$. The notation~$\ell_0$ and $u_0$ for the displacement and velocity datum, respectively, serves to emphasize that $d_{\SS}(0)$ and $d'_{\SS}(0)$ must coincide with the initial length and velocity of the fluid domain.

In accordance with the kinematic condition~\eqref{eq:kin1}, at the interface we identify the deformation of the fluid domain with the deformation of the structure domain. This leads to $d_{\SS}(t)=\ell(t)$. Kinematic condition~\eqref{eq:diric} implies $u_{\FF}(t)=d'_{\SS}(t)$. Equation~\eqref{eq:PS0} then provides a relation between the deformation and velocity of the structure at the interface, and the traction (in this case, pressure) exerted by the fluid on the structure: 
\begin{equation}
\label{eq:PS}
p_{\FF}(t,\ell(t))=\rho_{\FF}d_{\SS}(t)d''_{\SS}(t)-\kappa_{\textsc{f}}{}d_{\SS}'(t)
\end{equation}
Equation~\eqref{eq:PS} provides a map $d_{\SS}|_{\Gamma}\mapsto{}p_{\FF}|_{\Gamma}$. Such a map is generally referred to as a {\it displacement-to-pressure\/} or {\it Poincar\'e--Steklov\/} (PS) operator. By means of the PS operator, the fluid subsystem can formally be eliminated from the equation of motion for the structure, by replacing $p_{\FF}$ in the right-hand side of~\eqref{eq:massspring} by the right-hand side of~\eqref{eq:PS}. In practical applications, however, deriving a closed-form expression for the PS operator is prohibitively complicated.

The first term on the right-hand side of~\eqref{eq:PS} represents a force that is 
exerted by the fluid on the piston, that is proportional to the acceleration of the piston. The term therefore resembles an inertial force, and the multiplicative factor $\rho_{\FF}d_{\SS}(t)$ can be conceived of as an added mass to the piston. This term and, by extension, all terms derived from it, will be referred as {\it (artificial) added mass\/} terms. The second term on the right-hand side of~\eqref{eq:PS}, which emanates from the damping of the permeable cover, represents a force proportional to the velocity of the piston. 
This second term therefore resembles an auxiliary damping force and, accordingly, this term and all terms derived from it will be referred to as {\it (artificial) added damping\/} terms.

In the context of the above leaky-piston problem, subiteration with DN coupling can be condensed into the following sequence of initial-value problems: given an initial approximation $d_{\SS,0}:(0,t_{\fin})\to\IR$, solve
\begin{equation}
\label{eq:pistonsubit}
\left.
\begin{aligned}
m_{\SS}d''_{\SS,k}+\kappa_{\SS}d_{\SS,k}&=\rho_{\FF}d_{\SS,k-1}d_{\SS,k-1}''-\kappa_{\textsc{f}}{}d_{\SS,k-1}'\quad\text{in }(0,t_{\textsc{fin}})
\\
d_{\SS,k}(0)&=\ell_0
\\
d'_{\SS,k}(0)&=u_0
\end{aligned}
\right\}
\quad\text{for }k=1,2,\ldots
\end{equation}
In numerical solution procedures for FSI, the subiteration procedure is typically applied per time step and, accordingly, $t_{\fin}$ in~\eqref{eq:pistonsubit} corresponds to the time-step size, ${\tau}$. Moreover, the initial approximation~$d_{\SS,0}$ is generally obtained by extrapolation of the solution from a previous time step. This implies that $d_{\SS,k}=\ell_0+O({\tau})$ as ${\tau}\to+0$. Therefore, at the expense of an error $O({\tau})$, the relation between $d_{\SS,k-1}$ and $d_{\SS,k}$ in~\eqref{eq:pistonsubit} can be linearized, by replacing $d_{\SS,k-1}$ in the first term on the right-hand side by~$\ell_0$. To facilitate the presentation, we henceforth additionally assume that  $d'_{\SS,k}(0)=u_0$ for $k=0,1,2,\ldots$. By virtue of the initial conditions in~\eqref{eq:pistonsubit}, the assumption evidently holds for $k=1,2,\ldots$. For $k=0$, the assumption is satisfied if the extrapolation from the previous time step is at least second-order accurate. Denoting by~$d_{\SS}$ a fixed point of~\eqref{eq:pistonsubit}, i.e. a solution to the piston deformation in the FSI problem in the time interval under consideration, and by $\varepsilon_{k}=d_{\SS,k}-d_{\SS}$ the error between the iterative approximation after~$k$ iterations and the actual solution, and ignoring terms $O({\tau})$, it follows from~\eqref{eq:pistonsubit} that:
\begin{equation}
\label{eq:pistonsubiterr}
\left.
\begin{aligned}
m_{\SS}\varepsilon''_k+\kappa_{\SS}\varepsilon_k&=\rho_{\FF}\ell_0\varepsilon_{k-1}''-\kappa_{\textsc{f}}\varepsilon_{k-1}'\quad\text{in }(0,{\tau})
\\
\varepsilon_{k}(0)&=0
\\
\varepsilon'_{k}(0)&=0
\end{aligned}
\right\}
\quad\text{for }k=1,2,\ldots
\end{equation}
To condense the analysis, we define:
\begin{equation}
\label{eq:nondim}
\varepsilon_{{\tau},k}(s)=
\varepsilon_{k}({\tau}s),
\quad
\omega={\tau}\sqrt{\kappa_{\SS}/m_{\SS}}, 
\quad
\alpha_{\mathrm{m}}=\rho_{\FF}\ell_0/m_{\SS},
\quad
\alpha_{\mathrm{d}}={\tau}\kappa_{\textsc{f}}/m_{\SS}
\end{equation}
and recast~\eqref{eq:pistonsubiterr} into
\begin{equation}
\label{eq:pistonsubiterr0}
\left.
\begin{aligned}
\varepsilon''_{{\tau},k}+\omega^2\varepsilon_{{\tau},k}&=\alpha_{\mathrm{m}}\varepsilon_{{\tau},k-1}''-\alpha_{\mathrm{d}}\varepsilon_{{\tau},k-1}'\quad\text{in }(0,1)
\\
\varepsilon_{{\tau},k}(0)&=0
\\
\varepsilon'_{{\tau},k}(0)&=0
\end{aligned}
\right\}
\quad\text{for }k=1,2,\ldots
\end{equation}
The initial-value problem~\eqref{eq:pistonsubiterr0} can be solved (e.g. by means of Laplace transform) to obtain:
\begin{equation}
\label{eq:recursive}
\varepsilon_{{\tau},k}
=
\alpha_{\mathrm{m}}
\mathcal{L}_{\mathrm{m}}
\varepsilon_{{\tau},k-1}
+
\alpha_{\mathrm{d}}\mathcal{L}_{\mathrm{d}}
\varepsilon_{{\tau},k-1}
\end{equation}
where the linear operators $\mathcal{L}_{\mathrm{m}}$ and~$\mathcal{L}_{\mathrm{d}}$ are defined by
\begin{subequations}
\label{eq:LmLd}
\begin{align}
[\mathcal{L}_{\mathrm{m}}\varepsilon](s)
&=
\varepsilon(s)-\int\nolimits_0^s\omega\sin\big(\omega(s-\zeta)\big)\,\varepsilon(\zeta)\,d\zeta
\label{eq:Lm}
\\
[\mathcal{L}_{\mathrm{d}}\varepsilon](s)
&=
-\int\nolimits_0^s\cos\big(\omega(s-\zeta)\big)\,\varepsilon(\zeta)\,d\zeta    
\label{eq:Ld}
\end{align}
\end{subequations}
The operators~$\mathcal{L}_{\mathrm{m}}$ and~$\mathcal{L}_{\mathrm{m}}$ 
pertain to the added-mass and added-damping effects of the fluid, respectively. One can verify that the operators~$\mathcal{L}_{\mathrm{m}}$ and~$\mathcal{L}_{\mathrm{d}}$ commute. By virtue of this commutativity, it follows from the binomial identity that
\begin{equation}
\label{eq:binom}    
\varepsilon_{\tau,k}
=
\big(\alpha_{\mathrm{m}}
\mathcal{L}_{\mathrm{m}}
+
\alpha_{\mathrm{d}}
\mathcal{L}_{\mathrm{d}}\big)^k
\varepsilon_{\tau,0}
=
\bigg(\sum\nolimits_{l=0}^k{k \choose l}
\big(\alpha_{\mathrm{m}}\mathcal{L}_{\mathrm{m}}\big)^{l}
\big(\alpha_{\mathrm{d}}\mathcal{L}_{\mathrm{d}}\big)^{k-l}
\bigg)
\varepsilon_{\tau,0}
\end{equation}
Therefore, in order to assess the convergence properties of the recursive relation in~\eqref{eq:recursive}, it suffices to consider powers of the operators~$\alpha_{\mathrm{m}}\mathcal{L}_{\mathrm{m}}$ and~$\alpha_{\mathrm{d}}\mathcal{L}_{\mathrm{d}}$.

\subsection{Artificial added-damping effect}
\label{sec:artificialaddeddamping}
In this section, we examine the properties of the artificial added-damping operator,~$\alpha_{\mathrm{d}}\mathcal{L}_{\mathrm{d}}$. We first consider the scaling of the added-damping operator with the time step, $\tau$. The operator~$\mathcal{L}_{\mathrm{d}}$ in~\eqref{eq:Ld} depends on the time step
via the dependence of $\omega$ in~\eqref{eq:nondim} on~$\tau$. However, in numerical approximation methods, one is generally interested in small time steps, i.e. in the limit $\tau\to{}+0$. In this limit, the cosine term in~\eqref{eq:Ld} approaches unity and, accordingly $\|\mathcal{L}_{\mathrm{d}}\|=O(1)$ as~$\tau\to{}+0$.
For the coefficient~$\alpha_{\mathrm{d}}$ according to~\eqref{eq:nondim}, it holds that $\alpha_{\mathrm{d}}\propto\tau$. Therefore, $\|\alpha_{\mathrm{d}}\mathcal{L}_{\mathrm{d}}\|\propto\tau$ as~$\tau\to{}+0$, which implies that for sufficiently small time steps, the added-damping effect decreases proportional to the time step if the time step is reduced. In the limit $\tau\to+0$, the added-damping effect is essentially independent of the stiffness, $\kappa_{\SS}$. Regarding the dependence of the added-damping operator on the resistance coefficient, $\kappa_{\textsc{f}}$, and the structural mass, $m_{\SS}$, one can infer from~\eqref{eq:nondim} and~\eqref{eq:Ld} that $\|\alpha_{\mathrm{d}}\mathcal{L}_{\mathrm{d}}\|\propto\kappa_{\textsc{f}}$, uniformly, and $\|\alpha_{\mathrm{d}}\mathcal{L}_{\mathrm{d}}\|\propto{}m_{\SS}^{-1}$ in the limit $m_{\SS}\to\infty$. 

Next, we regard the specific properties of~$\mathcal{L}_{\mathrm{d}}$. The operator~$\mathcal{L}_{\mathrm{d}}~$ corresponds to an integral operator. The kernel in the integral operator and, hence, the operator itself are of Volterra type; see~\cite[Sec.3.2.3]{Kato:1984lo}. Set on a suitable domain, the operator is {\it compact\/}. In addition, it is {\it quasi-nilpotent\/}, which implies that its spectral radius 
$\operatorname{spr}(\mathcal{L}_{\mathrm{d}})=\lim_{k\to\infty}\|\mathcal{L}_{\mathrm{d}}^k\|^{1/k}$ vanishes and, in particular, it has no non-zero eigenvalues. The latter implies that~$\mathcal{L}_{\mathrm{d}}$  does not admit a spectral representation. Because every compact normal operator admits a spectral representation, it follows that~$\mathcal{L}_{\mathrm{d}}$ is {\it nonnormal\/}; see also~\cite{van2011partitioned,brummelen-borst}. A consequence of this nonnormality is that the sequence $\{\|(\alpha_{\mathrm{d}}\mathcal{L}_{\mathrm{d}})^k\|\}_{k\in\mathbb{N}}$ can display nonmonotonous convergence, i.e. divergence can precede asymptotic convergence, despite formal stability. In practice, this implies that the artificial added-damping effect can cause initial divergence of the iteration error in the subiteration method with DN coupling, despite formal asymptotic stability of the method. Such initial divergence of the iteration error is detrimental to the robustness of the iterative procedure. 

\begin{figure}
\begin{center}
\includegraphics[width=\textwidth]{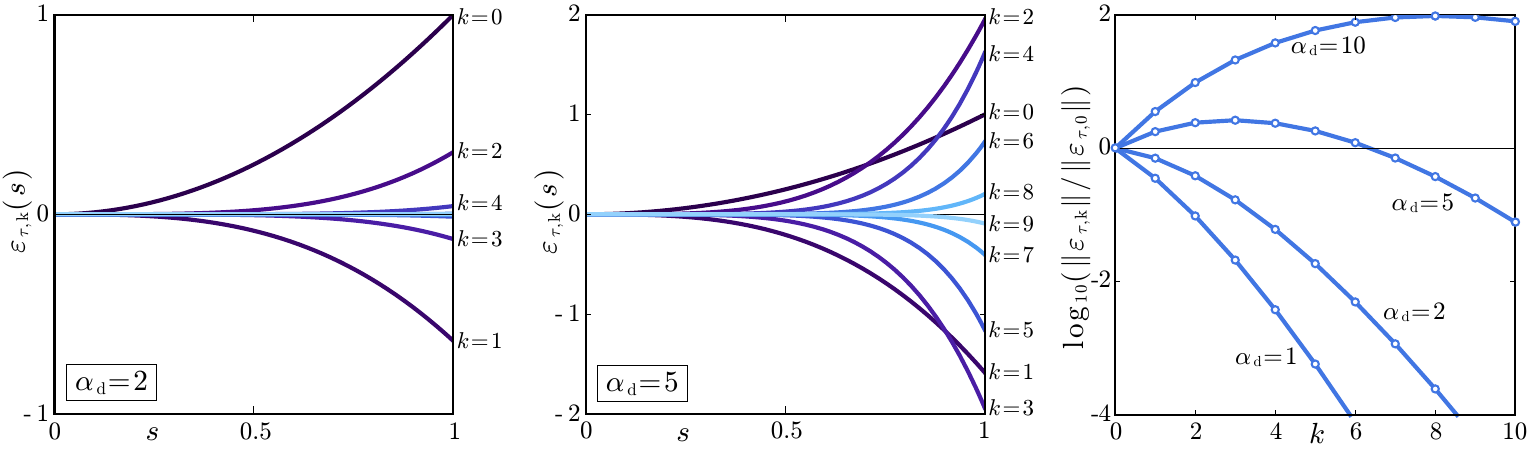}
\end{center}
\caption{Illustration of nonmonotonous convergence of subiteration with DN coupling for large added-damping effect:
Graph of
$\varepsilon_{\tau,k}=(\alpha_{\mathrm{d}}\mathcal{L}_{\mathrm{d}})^k\varepsilon_{\tau,0}$ for $\varepsilon_{\tau,0}(s)=s^2$ for $\alpha_{\mathrm{d}}=2$ ({\it left\/})
and $\alpha_{\mathrm{d}}=5$ ({\it center\/}), and $\log_{10}(\|\varepsilon_{\tau,k}\|_{H^1(0,1)}/\|\varepsilon_{\tau,0}\|_{H^1(0,1)})$ versus iteration counter~$k$ ({\it right\/}). 
\label{fig:adLd_convergence}}
\end{figure}

To illustrate the potential nonmonotonous convergence behavior caused by the artificial added-damping effect, Figure~\ref{fig:adLd_convergence} plots 
$\varepsilon_{\tau,k}=(\alpha_{\mathrm{d}}\mathcal{L}_{\mathrm{d}})^k\varepsilon_{\tau,0}$ with $\varepsilon_{\tau,0}(s)=s^2$ for $\alpha_{\mathrm{d}}=2$ ({\it left\/}) and~$\alpha_{\mathrm{d}}=5$ ({\it center\/}). In Figure~\ref{fig:adLd_convergence}, the parameter~$\omega$ in~\eqref{eq:Ld} has been arbitrarily set to~$\omega=1$. The right panel of Figure~\ref{fig:adLd_convergence} displays the ratio $\|\varepsilon_{\tau,k}\|_{H^1(0,1)}/\|\varepsilon_{\tau,0}\|_{H^1(0,1)}$ versus the iteration counter, $k$. One can observe that for sufficiently 
small~$\alpha_{\mathrm{d}}$, e.g. $\alpha_{\mathrm{d}}=2$, the iteration error $\varepsilon_{\tau,k}(s)$ coverges monotonously to zero as~$k$ increases
for all $s\in(0,1)$. Accordingly, the norm $\|\varepsilon_{\tau,k}\|_{H^1(0,1)}$ decays monotonously with~$k$.
However, for larger $\alpha_{\mathrm{d}}$, e.g. $\alpha_{\mathrm{d}}=5$, the convergence of $\varepsilon_{\tau,k}(s)$ is nonmonotonous for large~$s$. Consequently, the norm~$\|\varepsilon_{\tau,k}\|_{H^1(0,1)}$ can initially increase with~$k$, before asymptotic convergence occurs; see Figure~\ref{fig:adLd_convergence} ({\it right\/}). 

\subsection{Artificial added-mass effect}
\label{sec:artificialaddedmass}
We next consider the artificial-added-mass operator, $\alpha_{\mathrm{m}}\mathcal{L}_{\mathrm{m}}$. It is important to note that the operator~$\mathcal{L}_{\mathrm{m}}$ in~\eqref{eq:Lm} comprises two parts, viz. the identity operator and an integral operator. The integral operator depends on the time step via~$\omega$. In the limit $\tau\to+0$, the integral operator scales as $O(\tau^2)$, while the identity is clearly independent of~$\tau$. Hence, for sufficiently small time steps, the integral-operator part is negligible. Because the identity operator is evidently normal, it holds that $\|\alpha_{\mathrm{m}}\mathcal{L}_{\mathrm{m}}\|=\operatorname{spr}(\alpha_{\mathrm{m}}\mathcal{L}_{\mathrm{m}})=\alpha_{\mathrm{m}}$ as $\tau\to{}+0$. Because $\alpha_{\mathrm{m}}=\rho_{\FF}\ell_0/m_{\SS}$ is independent of~$\tau$, the artificial-added-mass effect is essentially independent of the time step,~$\tau$. This confirms previous findings on the added-mass effect of incompressible flows~\cite{causin2005added,van2009added,Forster:2007aa}. A further consequence is that if the ratio of the artificial added mass of the fluid to the mass of the structure, $\rho_{\FF}\ell_0/m_{\SS}$, exceeds one, then the subiteration method with DN coupling is unstable, independent of the time step. In such cases it is necessary to revert to stabilization/acceleration procedures, e.g. under-relaxation or quasi-Newton methods; see also Section~\ref{introduction}.

\section{Numerical examples}
\label{sec:num-ex}
This section presents numerical experiments for a representative model FSI problem with an incompressible fluid in a nearly-closed domain, with the main purpose of assessing to what extend the conclusions derived for the simple leaky-piston problem carry over to more complex FSI problems.

\subsection{Leaky-balloon-inflation test case}
\label{sec:Leakyballoon}
\begin{figure}
\centerline{\includegraphics[width=0.6\textwidth]{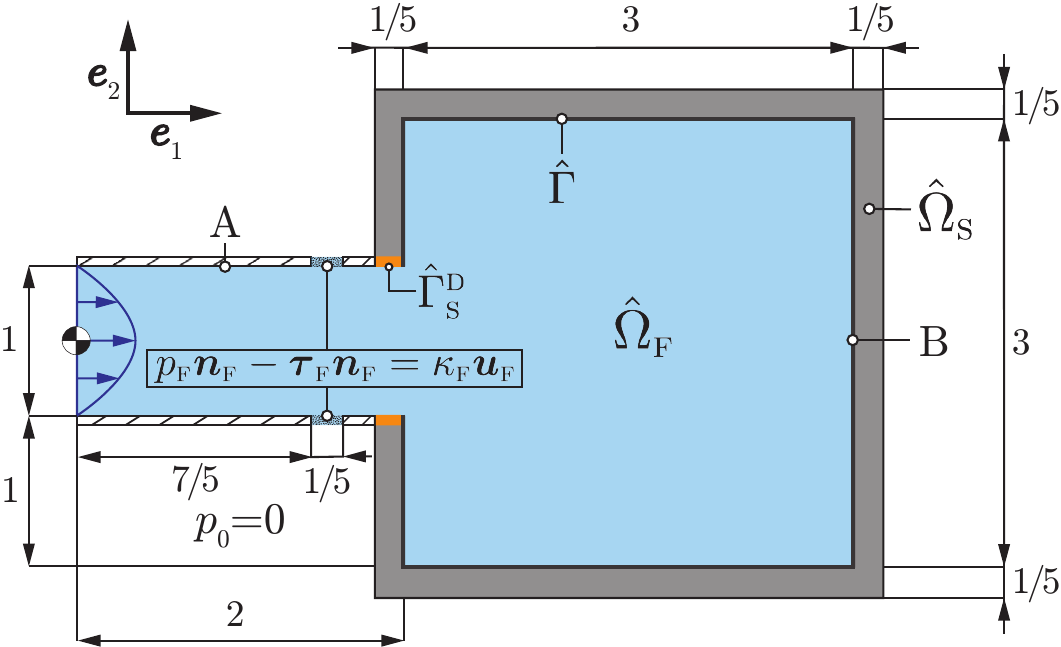}}
\caption{Illustration of the {\it leaky-balloon\/} inflation test case. The fluid domain consists of a $2\times 1$ inlet channel connected to a $3 \times 3$ square (blue). The square part is surrounded by a hyperelastic solid with thickness $1/5$ (gray). The top and bottom boundaries of the inlet channel comprise a porous region, with resistance $\kappa_{\textsc{f}}$. Point A with coordinates $(1,1/2)$ is a reference point to monitor the pressure. Point B with coordinates $(5,0)$ is a reference point to monitor displacement. (All lengths in [m])
\label{fig:LeakyBalloon}}
\end{figure}

We consider a modified version of the balloon-inflation FSI case, which has been proposed in~\cite{Kuttler} in the context of the incompressibility dilemma; see the illustration in Figure~\ref{fig:LeakyBalloon}. The main configurational modification comprises a small permeable section in the otherwise impermeable channel that connects the balloon to the inlet. For high resistances in the permeable section, the fluid domain is nearly closed. Formally, the closed scenario is recovered in the limit as the flow resistance passes to infinity.

The initial configuration of the fluid domain consists of a square 
$\hat{\Omega}^{\textrm{sq}}_{\textsc{f}}=(2,5)\times(-3/2,3/2)$
connected to an inflow boundary~$\Gamma_{\textsc{f}}^{\textsc{i}}=\{-2\}\times(-1/2,1/2)$ by a channel
$\hat{\Omega}^{\textrm{ch}}_{\textsc{f}}=(0,2)\times(-1/2,1/2)$:
\begin{equation*}
\hat{\Omega}_{\textsc{f}}
=
\operatorname{int}\big(
\operatorname{cl}\big(
\hat{\Omega}^{\textrm{sq}}_{\textsc{f}}
\big)
\cup
\operatorname{cl}\big(
\hat{\Omega}^{\textrm{ch}}_{\textsc{f}}
\big)
\big)
\end{equation*}
with $\operatorname{int}(\cdot)$ and~$\operatorname{cl}(\cdot)$ as interior and closure, respectively. The structure subsystem occupies the domain 
\begin{equation*}
\hat{\Omega}_{\textsc{s}}=(9/5,26/5)\times(-17/10,17/10)\setminus\hat{\Omega}_{\textsc{f}},    
\end{equation*}
forming a $1/5$\nobreakdash-thick layer around the boundary of the square part of the fluid domain, except the part connected to the channel; see Figure~\ref{fig:LeakyBalloon}. The initial configuration of the deformable fluid-structure interface hence corresponds to $\hat{\Gamma}=\partial\hat{\Omega}^{\textrm{sq}}_{\textsc{f}}\setminus\partial\hat{\Omega}^{\textrm{ch}}_{\textsc{f}}$.

The lateral boundaries of the channel, 
$\Gamma_{\textsc{f}}^{\textsc{w}}=(-2,0)\times\{\pm1/2\}\setminus\Gamma_{\textsc{f}}^{\textsc{r}}$, correspond to no-slip walls and the fluid velocity is subject to homogeneous Dirichlet boundary conditions of the form~\eqref{eq:bcFD} at these boundaries, except for two perforated sections $\Gamma_{\textsc{f}}^{\textsc{r}}=(7/5,8/5)\times\{\pm1/2\}$ in the top and bottom boundaries which are furnished with a Robin boundary, from which the fluid can exit the domain subject to a flow resistance~$\kappa_{\textsc{f}}$; see Equation~
\eqref{eq:bcFR}. The perforations have been introduce in both the top and bottom boundaries to retain symmetry of the configuration with respect to the $\boldsymbol{e}_1$-axis.

At the inflow boundary, a time-dependent parabolic profile is imposed by means of a Dirichlet condition:
\begin{equation}
\label{eq:inflow}
\boldsymbol{u}_{\textsc{f}}(t,x_1,x_2)
=
\theta(t)\,U(x_2)\,\boldsymbol{e}_1
\qquad\text{in }(0,t_{\textsc{fin}})\times\Gamma_{\textsc{f}}^{\textsc{i}}
\end{equation}
where $U(x_2)$ is a parabolic function such that $U(0)=1$ and $U(\pm1/2)=0$, and $\theta(t)$ is a time-dependent function:
\begin{equation}
\label{eq:theta}
\theta(t)
=
\begin{cases}
\frac{1}{2}-\frac{1}{2}\cos(\pi{}t)&\quad{}t\in[0,1)
\\
1&\quad{}t\in[1,\infty),
\end{cases}
\end{equation}
The structure is rigidly fixed at the edges adjacent to the channel by means of a Dirichlet condition:
\begin{equation*}
\boldsymbol{d}_{\textsc{s}}(t,\cdot)=(\cdot)
\qquad\text{on }\hat{\Gamma}_{\textsc{s}}^{\textsc{d}},\:{}t\in(0,t_{\textsc{fin}})
\end{equation*}
with $\hat{\Gamma}_{\textsc{s}}^{\textsc{d}}=\partial\hat{\Omega}_{\textsc{s}}\cap\partial\hat{\Omega}_{\textsc{s}}^{\text{ch}}$. The structure is traction free at its external boundary and, accordingly, satisfies a homogeneous Neumann condition of the form~\eqref{eq:bc-solidN} on 
$\hat{\Gamma}_{\textsc{s}}^{\textsc{n}}=\partial\hat{\Omega}_{\textsc{s}}
\setminus\partial\hat{\Omega}_{\textsc{f}}$.

As quantities of interest, we monitor the pressure $p_{\textsc{A}}$ at the midpoint of the top boundary of the channel, and the horizontal displacement $u_{\textsc{B}}$ at the mid point of the frontal edge of the fluid domain; see Figure~\ref{fig:adLd_convergence}.

\subsection{Leaky-to-closed convergence}
\label{sec:L2C}
We first consider the effect of the leaky sections on the behavior of the balloon-inflation process. To this end, we compute the evolution of the leaky-balloon-inflation problem,
equipped with homogeneous initial conditions, for various flow-resistance parameters, and compare the results with those of the fully closed configuration. Because a DN-partitioned solution strategy will fail for the fully closed case and for large values of the resistance parameter, we conduct the investigations by means of a monolithic solution strategy implemented in the open-source software Nutils~\cite{nutils70}. For validation purposes, the results for the fully closed case are compared to those obtained by a DN coupling scheme with a Lagrange multiplier to enforce the compatibility condition.

For this test case, the structure is modeled as a hyperelastic material with St. Venant--Kirchhoff constitutive behavior, characterized by the strain-energy-density function
\begin{equation}
\label{eq:StVenantKirchhoff}
W(\boldsymbol{E}_{\textsc{s}}) = G_{\textsc{s}}\operatorname{tr}(\boldsymbol{E}_{\textsc{s}}^2)+\tfrac{1}{2}\lambda_{\textsc{s}}(\operatorname{tr}\boldsymbol{E}_{\textsc{s}})^2
\end{equation}
with $G_{\SS}=241379$\,Pa as the shear modulus and $\lambda_{\SS}=217241$\,Pa as the first Lam\'e parameter, such that the Young's modulus $E_{\textsc{s}}=G_{\textsc{s}}(3\lambda+2G)/(\lambda_{\textsc{s}}+G_{\textsc{s}})=7\times{}10^5\,\text{Pa}$ and the Poisson ratio $\nu_{\textsc{s}}=\tfrac{1}{2}\lambda_{\textsc{s}}/(\lambda_{\textsc{s}}+G_{\textsc{s}})=0.45$, in accordance with the setup in~\cite{Kuttler,Bogaers2015}; see also Table~\ref{tab:TC1parameters} (TC1).
\begin{table}
\centering
\caption{Fluid and solid material properties for the leaky-balloon inflation test cases: fluid density $\rho_{\textsc{f}}$, fluid dynamic viscosity $\mu_\textsc{f}$,
resistance parameter $\kappa_{\FF}$,
solid density $\rho_{\textsc{s}}$,  Young's modulus $E_{\textsc{s}}$, Poisson ratio $\nu_{\textsc{s}}$, and time step ${\tau}$. Entries marked as $\ast$ are varied in the test case.\label{tab:TC1parameters}}
 \begin{tabular}{c c c c c c c c} 
 \hline
TC\# & 
$\rho_{\textsc{f}}$~[kg/m$^3$] & 
$\mu_{\textsc{f}}$~[Pas] & 
$\kappa_{\textsc{f}}$~[kg/m$^2$s] &
$\rho_{\textsc{s}}$~[kg/m$^3$] & $E$ [Pa] & 
$\nu_{\textsc{s}}$ & 
${\tau}$ [s]\\ 
 \hline
1 & $1.1$ & $0.1606$ & $\ast$ & $1\times10^3$          & $7\times 10^5$ & $0.45$ & $5\times{}10^{-3}$\\
2 & $10^{-2}$ & $0.1606$ & $\ast$ & $5\times{}10^3$  & $7\times 10^5$ & $0.45$ & $10^{-2}$\\
3 & $10^{-2}$ & $0.1606$ & $5\times{}10^3$ & $5\times{}10^3$  & $7\times 10^5$ & $0.45$ & $\ast$ \\
4 & $10^{-2}$ & $0.1606$ & $5\times10^3$ & $\ast$  & $7\times 10^5$ & $0.45$ & $10^{-2}$ \\
5 & $\ast$ & $0.1606$ & $5\times10^3$ & $5\times{}10^3$  & $7\times 10^5$ & $0.45$ & $10^{-2}$ \\
\hline
 \end{tabular}
\end{table}

In addition to the perforations in the top and bottom boundaries, the presented test case deviates from that in~\cite{Kuttler} by the time-dependence of the inflow condition in~\eqref{eq:theta} and the stress-strain relation of the solid. While~\cite{Kuttler} applies a  ramp-up function $\sin(\pi{}t/2)$ in the interval $t\in[0,1]$, we opt to use $\frac{1}{2}-\frac{1}{2}\cos(\pi{}t)$, analogous to~\cite{Bogaers2015}, to enhance the smoothness of the solution in time. We apply a St.Venant--Kirchhoff consitutive relation for the solid, in accordance with~\eqref{eq:StVenantKirchhoff}, instead of the neo-Hookean constitutive relation in~\cite{Kuttler}, because the St.Venant--Kirchhoff relation is uniquely defined as a constitutive model, while the neo-Hookean models represent a constitutive class which requires further specification. In addition, the reference results provided to us, obtained with DN coupling with a Lagrange multiplier for the volume constraint, are based on the St.Venant--Kirchhoff stress-strain relation.

For completeness, we report some aspects of the numerical setup. We consider numerical approximations based on a finite-element approximation in the spatial dependence, and an implicit Euler approximation in the temporal dependence. The domain deformation in the fluid domain is represented by means of a pseudo solid with Jacobian-based stiffening.  
The resulting sequence of non-linear algebraic systems, conforming to~\eqref{eq:coupled}, is solved by means of a standard Newton procedure, in which the linear tangent problems are solved by means of a direct solver. The solution procedure hence corresponds to a monolithic method. For the approximation in the spatial dependence, we apply piecewise quadratic finite-element approximations for deformation and velocity, and piecewise linear finite-element approximations for the fluid pressure. The incompressible Navier--Stokes equations are hence approximated by a standard $P_2-P_1$ Taylor--Hood velocity-pressure pair, while the deformations of the solid and pseudo-solid are approximated by $P_2$ finite elements. We consider finite-element meshes comprised of square element domains of size $1/20\times1/20$; see also Figure~\ref{fig:dispevol} below. In view of the symmetry of the configuration (see Figure~\ref{fig:LeakyBalloon}) with respect to the $\boldsymbol{e}_1$\nobreakdash-axis, we regard only one half of the configuration.

\begin{figure}
\centerline{%
\includegraphics[height=0.3\textwidth]{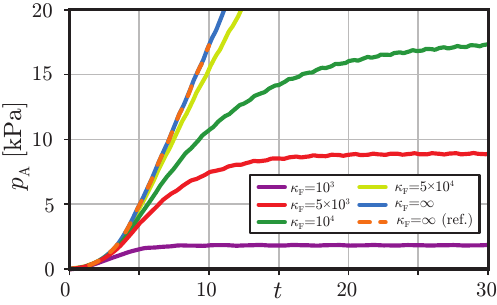}
\includegraphics[height=0.3\textwidth]{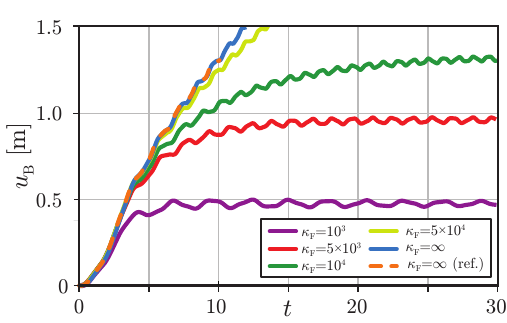}}
\caption{Pressure at gauge point A ({\it left\/}) and  horizontal displacement at gauge point B ({\it right\/}) for the fully closed scenario ($\kappa_{\FF}=\infty$) and for $\kappa_{\FF}\in\{10^3,5\times{}10^3,10^4,5\times10^4\}$ [kg/m$^2$s]. The reference result for the fully closed case has been obtained with DN coupling with a Lagrange multiplier.
\label{fig:LB_pd}}
\end{figure}

Figure~\ref{fig:LB_pd} plots the pressure at gauge point A ({\it left\/}) and the horizontal displacement at gauge point B ({\it right\/}) for the fully closed scenario and for $\kappa_{\FF}\in\{10^3,5\times{}10^3,10^4,5\times10^4\}$ [kg/m$^2$s]. One can observe that for finite resistance parameters, $\kappa_{\FF}$, the pressure and displacement initially increase, and ultimately settle into a steady regime of quasi-periodic fluctuations around a constant value. The quasi-periodic behavior corresponds to a so-called breathing motion of the balloon. The displacement oscillations are more pronounced at lower $\kappa_{\FF}$.
Figure~\ref{fig:LB_pd} moreover conveys that the pressure and displacement evolution for the leaky balloon, i.e. at finite resistance~$\kappa_{\FF}$, approach those corresponding to the closed balloon in the limit $\kappa_{\FF}\to\infty$. Hence, the closed-balloon case can indeed be viewed as the limit of the leaky-balloon case as $\kappa_{\FF}\to\infty$.
It is to be mentioned that the monolithic procedure is uniformly robust in $\kappa_{\FF}$, in the sense that the convergence behavior of the Newton procedure and of the underlying direct solver for the linear tangent problems are essentially independent of $\kappa_{\FF}$, and essentially identical to that of the fully closed case.

To further illustrate the dynamics of the leaky-balloon test case, Figure~\ref{fig:dispevol} displays the configuration of the leaky balloon at $t\in\{0,1,5,40\}$. The configuration at $t=0$ corresponds to the initial undeformed configuration.
The colors in the figure represent the magnitude of the fluid velocity. One can observe the deformation of the solid and, correspondingly, of the fluid domain as time progresses, due to
inflow at the inlet. The result at $t=1$ conveys that 
the leakage velocity is initially negligible. This is in agreement with the fact that the pressure is initially low
(see Figure~\ref{fig:LB_pd}) and, hence, the Robin condition~\eqref{eq:bcFR} insists that the leakage velocity is small as well. The result at $t=5$ shows that as time progresses, the leakage velocity increases, in accordance with the build up of pressure in the balloon. The velocity field at $t=40$ indicates that, ultimately, the leakage at the permeable sections balances the inflow, apart from the small volumetric variations caused by the breathing motion of the balloon.
\begin{figure}
\centerline{%
\includegraphics[width=1\textwidth]{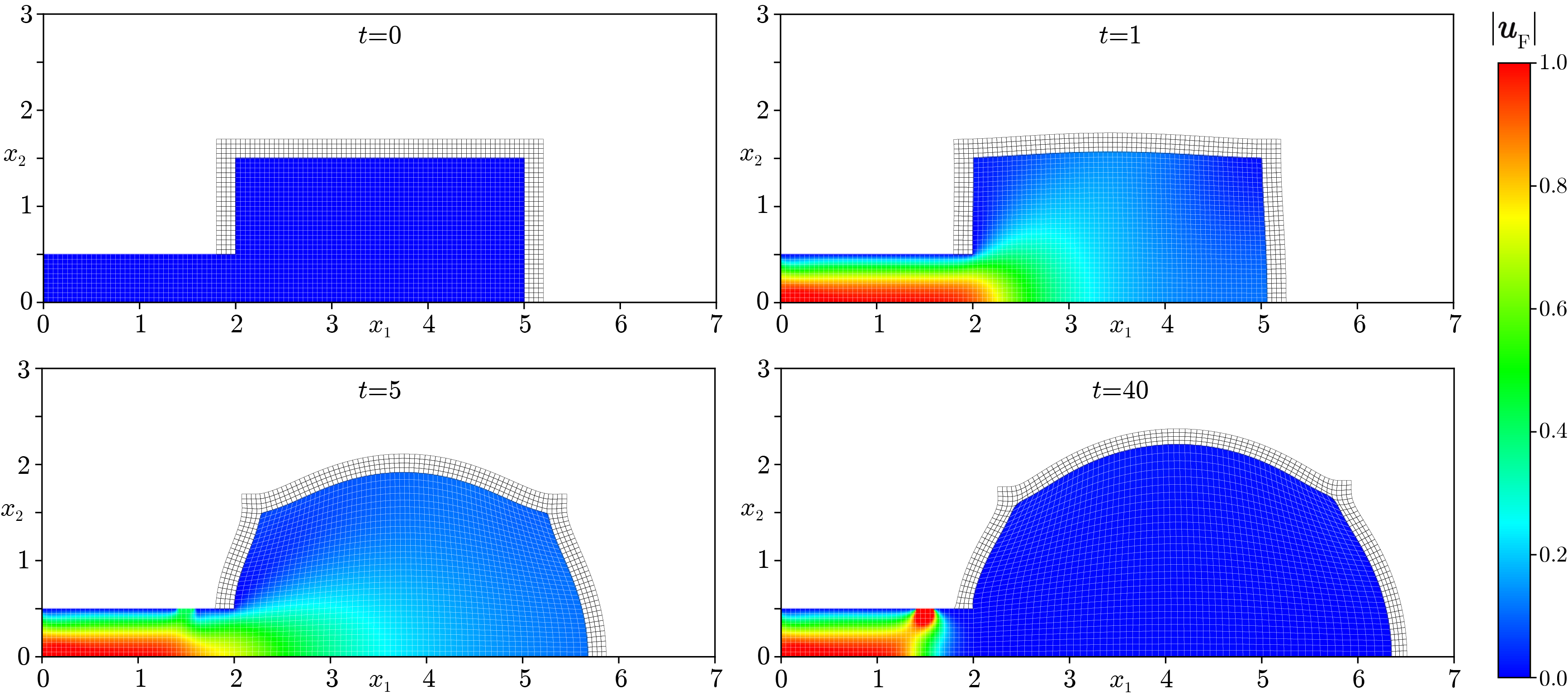}}
\caption{Evolution of the leaky-balloon configuration: deformation at $t\in\{0,1,5,40\}$. Colors indicate fluid velocity.
\label{fig:dispevol}}
\end{figure}

\subsection{Stability and convergence of DN coupling}
\label{sec:SCDN}
We next consider the convergence behavior of DN-coupled subiteration for the leaky-balloon inflation case, in relation to the main parameters of the problem. We conduct this investigation in the Ansys Fluent framework, composed of Ansys Fluent for the fluid subsystem, Ansys Mechanical for the solid subsystem, and ANSYS SystemCoupling to map and transfer data between the fluid and solid subsystems (ANSYS Inc., Canonsburg, PA, USA). Ansys Fluent applies a finite-volume approximation of an ALE formulation of the 
momentum equation~\eqref{eq:NS_mom} with a second-order upwind scheme for the discretization of the convective terms, and a second-order least-squares cell-based approximation of the diffusive fluxes. A first-order Backward Euler scheme is applied to discretize the momentum equation in the temporal dependence. For the fluid-mesh deformation, we apply simple diffusion-based smoothing; see Section~\ref{sec:moving_domain}. Ansys Mechanical applies a finite-element approximation of the solid subsystem in the spatial dependence. We opt to use quadratic elements. For the temporal discretization, a Hilber--Hughes--Taylor (HHT) $\alpha$ method is used. We consider an approximation in which both the fluid and the solid (in the reference configuration) are approximated with square elements of size $1/20\times1/20$.
The coupled FSI problem in each time step is solved by means of subiteration with DN coupling, provided with an initial estimate for the solution corresponding to the converged solution from the previous time step. 

We introduce some adjustments to the leaky-balloon test-case described above, to render it amenable to simulation by Ansys Fluent. Because Ansys Mechanical does not provide St.Venant--Kirchhoff constitutive behavior, we apply a Neo--Hookean constitutive relation instead, with the same Young's modulus and Poisson ratio as considered in Section~\ref{sec:L2C}; see also Table~\ref{tab:TC1parameters}. Since we regard the convergence behavior of DN coupling at small strains, we expect however that the results are insensitive to the specifics of the constitutive model. In particular, we regard the convergence behavior immediately after the inflow velocity~\eqref{eq:theta} has settled into its stationary regime at $t=1$\,s, and at this instant the deformations and strains are small; see Figure~\ref{fig:dispevol}. In addition, because Ansys Fluent is not equipped with a Robin-type boundary condition, in the numerical implementation, the Robin boundary condition at the permeable part of the inlet channel is represented by a porous region. This porous region is $1/5$ (units:~m) wide and $\xi=1/10$ thick, 
located at~$\Gamma_{\FF}^{\textsc{r}}$. The porous region is modeled by Darcy's equation, viz. $\nabla{}p_{\textsc{f}}=-(\mu_{\textsc{f}}/k_{\FF})\boldsymbol{u}_{\textsc{f}}$ with~$k_{\FF}$ as the permeability. Because the pressure in the porous layer at the inside and outside correspond to the fluid traction and $p_0\boldsymbol{n}$, respectively, the porous region acts as a Robin-type boundary condition for the fluid problem with resistance $\kappa_{\textsc{f}}=\xi\mu_{\textsc{f}}/k$.

\begin{figure}
\centerline{\includegraphics[width=0.5\textwidth]{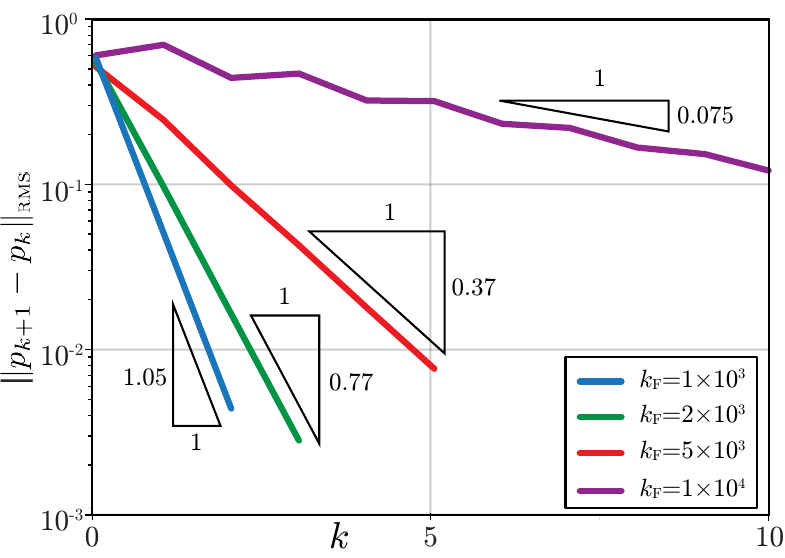}}
\caption{Dependence of the convergence behavior of subiteration on the flow resistance, $\kappa_{\FF}$: Norm of the pressure update $\|p_{k+1}-p_k\|_{\textsc{rms}}$ in the subiteration process versus iteration counter $k$ for TC2.}
\label{fig:kappa_F}
\end{figure}

We first examine the effect of flow resistance on the convergence behavior of subiteration with DN coupling. The parameters for this test case are presented in Table~\ref{tab:TC1parameters} (TC2). To reduce the added-mass effect relative to the added-damping effect, we consider a reduced fluid density of $\rho_{\FF}=10^{-2}$. Furthermore, to reduce both the added-mass and added-damping effects even further, we increase the solid density to $\rho_{\SS}=5\times{}10^3$. We then monitor the update of the 
pressure  that is transferred from the fluid to the solid in each iteration, at $t=1$. 
Figure~\ref{fig:kappa_F} presents the RMS value of the pressure update versus the number of iterations of the subiteration procedure with DN coupling, for $\kappa_{\FF}\in\{10^3,2\times{}10^3,5\times10^3,10^4\}$. The oscillations in the convergence behavior that occur for $\kappa_{\FF}=10^4$ can conjecturally be attributed to the fact that for large $\kappa_{\FF}$, the fluid pressure may still exhibit relatively large variations, even at low tolerances in the Ansys Fluent solver.
One can observe that the convergence rate of the subiteration process deteriorates as the resistance, $\kappa_{\FF}$, increases.  
The slopes of the curves, indicated by the triangles in Figure~\ref{fig:kappa_F}, are to close approximation related by
\begin{equation}
a^{\star}-a=\log_{10}(\kappa_{\FF}^{\star}/\kappa_{\FF})
\end{equation}
For instance, the slopes for $\kappa_{\FF}^{\star}=10^4$ and $\kappa_{\FF}=10^3$ satisfy $-0.075-(-1.05)\approx{}1=\log_{10}(\kappa_{\FF}^{\star}/\kappa_{\FF})$. The relation between the slopes of the curves indicates that the convergence rate for the leaky-balloon problem is proportional to $\kappa_{F}$, in agreement with the theory for the simple leaky-piston problem presented in Section~\ref{sec:artificialaddeddamping}, in particular, with the linear dependence of~$\alpha_{\mathrm{d}}$ in~\eqref{eq:nondim} on~$\kappa_{\FF}$. 

\begin{figure}
\centerline{\includegraphics[width=0.5\textwidth]{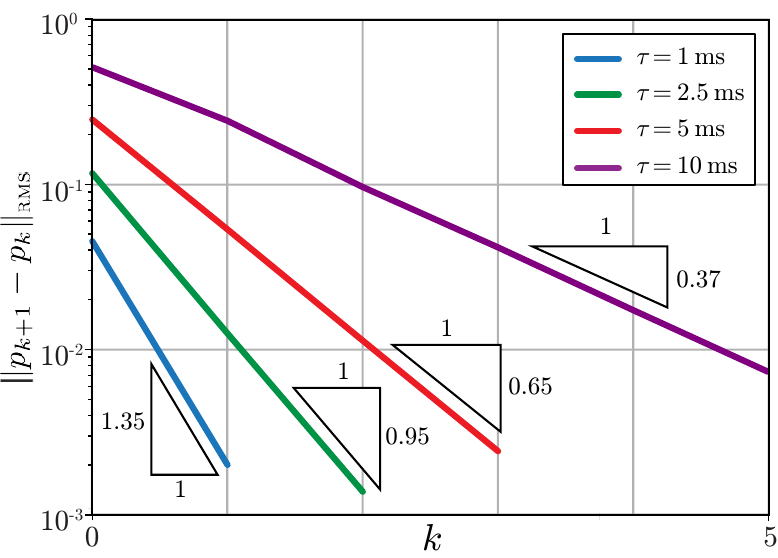}}
\caption{Dependence of the convergence behavior of subiteration on the time step, ${\tau}$: Norm of the pressure update $\|p_{k+1}-p_k\|_{\textsc{rms}}$ in the subiteration process versus iteration counter $k$ for TC3.}
\label{fig:tau}
\end{figure}

We next consider the dependence of the convergence behavior of the subiteration process on the time step. The setup is identical to that of TC2, except that the resistance $\kappa_{\FF}$ is fixed at $5\times{}10^3$ and the time step in the simulation is varied. Table~\ref{tab:TC1parameters} (TC3) summarizes the parameters. Figure~\ref{fig:tau} displays the magnitude of the pressure update versus the number of iterations of the subiteration procedure, for ${\tau}\in\{1,2.5,5,10\}$\,ms. One can observe that the initial update reduces as ${\tau}$ decreases. This can be attributed to the fact the initial estimate in the subiteration procedure corresponds to the converged solution of the previous time step. Accordingly, the error in the initial estimate behaves as~$O({\tau})$, which implies that the initial estimate becomes increasingly accurate as~${\tau}$ decreases. The results 
in Figure~\ref{fig:tau} convey that the convergence rate deteriorates as the time step increases. Considering the slopes of the curves in Figure~\ref{fig:tau}, by similar arguments as for TC2, one can infer that the convergence rate 
of the subiteration process is proportional to~${\tau}$. This is consistent with the theory for the leaky-piston problem in  
Section~\ref{sec:artificialaddeddamping}.

\begin{figure}
\centerline{\includegraphics[width=0.5\textwidth]{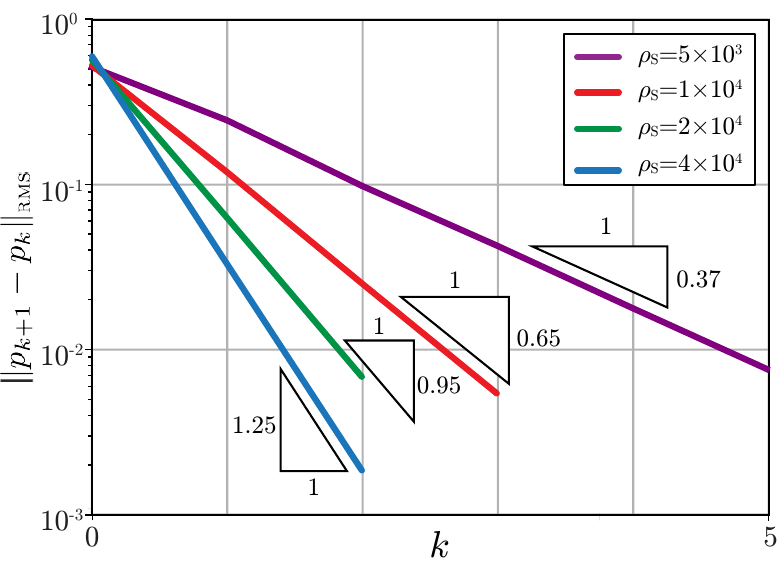}}
\caption{Dependence of the convergence behavior of subiteration on the solid density, $\rho_{\SS}$: Norm of the pressure update $\|p_{k+1}-p_k\|_{\textsc{rms}}$ in the subiteration process versus iteration counter $k$ for TC4.}
\label{fig:rho_s}
\end{figure}

Section~\ref{sec:artificialaddeddamping} conveys that the convergence behavior of the subiteration process improves as the solid density increases. To investigate the extension of this result to the leaky-balloon case, we consider a setup 
analogous to TC2, except that the resistance $\kappa_{\FF}$ is fixed at $5\times{}10^3$ and the solid density, $\rho_{\SS}$, is varied; see Table~\ref{tab:TC1parameters} (TC4). Figure~\ref{fig:rho_s} displays the magnitude of the pressure update versus the number of iterations for $\rho_{\SS}\in\{5\times{}10^3,10^4,2\times{}10^4,4\times{}10^4\}$\,kg/m$^3$. One can oberve from Figure~\ref{fig:rho_s} that the convergence behavior of the subteration process improves as $\rho_{\SS}$ increases. The slopes of the curves in Figure~\ref{fig:rho_s} indicate that the convergence rate of the subiteration procedure is inversely proportional to~$\rho_{\SS}$, which is commensurate with the dependence of~$\alpha_{\mathrm{d}}$ in~\eqref{eq:nondim} on~$\rho_{\SS}$.

\begin{figure}
\centerline{\includegraphics[width=0.5\textwidth]{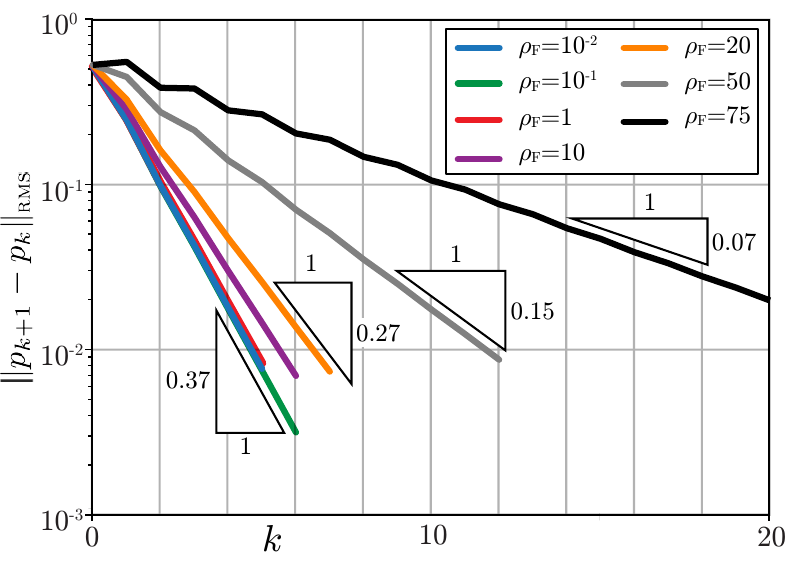}}
\caption{Dependence of the convergence behavior of subiteration on the fluid density, $\rho_{\FF}$: Norm of the pressure update $\|p_{k+1}-p_k\|_{\textsc{rms}}$ in the subiteration process versus iteration counter $k$ for TC5.}
\label{fig:rho_f}
\end{figure}

We finally examine the dependence of the convergence behavior of the subiteration process with DN coupling on the fluid density, $\rho_{\FF}$. To this purpose, we consider a setting similar to~TC2, but with fixed resistance $\kappa_{\FF}=5\times{}10^3$ and varying~$\rho_{\FF}$; see Table~\ref{tab:TC1parameters} (TC5).
Figure~\ref{fig:rho_f} presents the magnitude of the pressure update versus the number of iterations for 
$\rho_{\FF}\in\{10^{-2},10^{-1},1,10,20,50,75\}$\,kg/m$^3$. One can note that for low fluid densities, the convergence behavior of the subiteration process is essentially independent of~$\rho_{\FF}$. 
This is consistent with the results for the leaky-piston problem:
For small fluid densities and sufficiently large $\kappa_{\FF}$ and~${\tau}$, the added-mass effect is subordinate to the 
added-damping effect, and the added damping effect is independent of~$\rho_{\FF}$; cf. $\alpha_{\mathrm{d}}$ in~\eqref{eq:nondim}.
For larger~$\rho_{\FF}$, the added-mass effect becomes more prominent, which explains the deviation of the curves for 
$\rho_{\FF}\in\{10,20,50,75\}$ relative to those for~$\rho_{\FF}\in\{10^{-2},10^{-1},1\}$. The 
theory for the leaky-piston problem in  
Section~\ref{sec:artificialaddeddamping} indicates that the convergence rate of the DN-coupled subiteration process is proportional to~$\rho_{\FF}$ if the added-mass effect is dominant, i.e. for sufficiently large~$\rho_{\FF}$. 
The curves for $\rho_{\FF}\in\{20,50,75\}$ in Figure~\ref{fig:rho_f} corroborate this result: The   
deviation between the difference in the slopes of two curves, $a^{\star}-a$, and the logarithm of the ratio of the corresponding fluid densities, $\log_{10}(\rho_{\FF}^*/\rho_{\FF})$, decreases as the fluid density increases.
Indeed, for $\rho_{\FF}^*=50$ and $\rho_{\FF}=20$, the deviation is $0.28$, while for $\rho_{\FF}^*=70$ and $\rho_{\FF}=50$ the deviation has reduced to $0.07$.

Summarizing the results for TC2\nobreakdash-5, we find that the dependence of the convergence behavior of the subiteration process with DN coupling on the main parameters of the FSI problem carries over from the simple leaky-piston problem to the leaky-balloon problem. For settings in which the added-damping effect dominates over the added-mass effect, the convergence rate is proportional 
to~$\tau\kappa_{\FF}/\rho_{\SS}$, and essentially independent of~$\rho_{\FF}$. For settings in which the added-damping effect is subordinate to the added-mass effect, the convergence rate is proportional to~$\rho_{\FF}$. At variance with the analysis of the leaky-piston problem, TC2\nobreakdash-5 do not display non-monotonous convergence behavior. We conjecture that such non-monotonous behavior can occur for larger values of~$\alpha_{\mathrm{d}}$. Increasing $\tau,\kappa_{\FF}$ and $\rho_{\SS}^{-1}$ beyond the values considered in TC2\nobreakdash-5 however leads to non-robustness of the Ansys Fluent solution procedure.

\section{Conclusion}
\label{sec:conclusion}
Subiteration with a Dirichlet--Neumann partitioning of the interface conditions is the standard approach for solving fluid-structure-interaction problems, by virtue of the fact that this approach retains modularity, thus enabling reuse of software, and translates into canonical boundary conditions for the fluid and structure subsystems. Subiteration with DN coupling however leads to the so-called incompressibility dilemma if the fluid is incompressible and the fluid domain is closed, in the sense that 
it is furnished with Dirichlet boundary conditions on the part of its boundary complementary to the fluid-structure interface. In practice, however, for instance in FSI analyses of valve systems, one generally observes that the subiteration procedure also fails for nearly closed configurations.

Motivated by the non-robustness of subiteration with DN coupling for nearly closed FSI problems, this work has investigated the convergence behavior of subiteration for nearly-closed scenarios, i.e. if the fluid domain is furnished with Dirichlet conditions except for a (small) permeable part of the boundary where a Robin-type condition holds, with a large flow resistance, $\kappa_{\FF}$. We established that for nearly closed fluid domains, volume-rate deviations in the fluid domain lead to pressure variations that are proportional to the flow resistance, and inversely proportional to the area of the Robin boundary, portending non-robustness of the DN scheme for nearly-closed FSI problems.

Based on a simple model problem, viz. the leaky-piston problem, we inferred that the non-robustness of subiteration with DN coupling for nearly closed problems can be attributed to a so-called {\it added-damping effect\/}. This added-damping effect is associated with a nonnormal operator. The nonnormality of the added-damping operator can lead to nonmonontonous convergence behavior, i.e. transient divergence can occur before asymptotic convergence sets in. Such transient divergence severely degrades the robustness of the subiteration process. The analysis of the leaky-piston problem moreover conveys that the norm of the added-damping operator is proportional to $\tau\kappa_{\FF}/\rho_{\SS}$ with $\tau$ as the time step and $\rho_{\SS}$ as the solid density. This indicates that, indeed, the convergence behavior of the subiteration process deteriorates as the flow resistance increases. However, the convergence behavior can be controlled by adapting the time step in the numerical procedure.

By means of numerical experiments for the leaky-balloon problem, we investigated the extension of the results for the leaky-piston problem to a more sophisticated nearly closed FSI problem. By means of a monolithic solution procedure, we established that the fully closed case is recovered in the limit as $\kappa_{\FF}\to\infty$, noting that the convergence behavior of the monolithic solver is essentially independent of~$\kappa_{\FF}$. For the DN coupled approach, we observed that the convergence rate is indeed proportional to $\tau\kappa_{\FF}/\rho_{\SS}$, corroborating the outcome of the analysis of the leaky-piston problem.

An important corollary of this investigation is that the standard DN-coupled subiteration scheme is inherently unsuitable for fluid-structure-interaction problems involving an incompressible fluid on a nearly-closed domain, or on a domain that continuously transforms from an open to a closed configuration, e.g. in valves.

\section*{Acknowledgments}
The authors are grateful to Prof. Miguel A. Fern\'andez for providing the reference results for the fully closed balloon test case, reported in Figure~\ref{fig:LB_pd}. The authors gratefully acknowledge the support of the European Union Horizon 2020 research and innovation program by funding the Ph.D. fellowship of Ahmed Aissa Berraies, which is part of the Marie Sk\l{}odowska-Curie ITN-EJD ProTechTion actions, with Grant Number 764636. 
The authors acknowledge also partial financial support from the Ministry of Enterprise and Made in Italy and Lombardy Region through the project
"PRotesi innOvaTivE per applicazioni vaScolari ed ortopedIChe e mediante Additive Manufacturing" - CUP : B19J22002460005, on the finance  Asse I, Azione 1.1.3 PON Businesses and Competitiveness 2014 - 2020. The authors are also grateful for the computing time and power provided by SurfSara center through access to Snellius Supercomputer capabilities, which were essential for performing the simulations. Thanks are also due to Ansys Inc. for providing the academic and HPC licenses required to conduct the present research.





\subsection*{Conflict of interest}

The authors declare that they have no known competing interests or personal relationships that could have appeared to influence the work reported in this paper.

\bibliography{wileyNJD-AMA}%
\bibliographystyle{plain}
\end{document}